\documentclass[prd,aps,nofootinbib,showpacs]{revtex4}

\usepackage{amsmath,amsfonts,amssymb,amsthm}           
\usepackage{dsfont}                                    
\usepackage{graphicx, epsfig}                          
\usepackage{multirow}

\usepackage{epstopdf}                                 

\usepackage{color}
\usepackage{verbatim}
\usepackage{hyperref}

\newcommand{\idMatrix}{\mathds{1}}
\newcommand{\pd}{\partial}
\newcommand{\ii}{\mathrm{i}}
\newcommand{\const}{\mathrm{const}}
\newcommand{\diff}{\mathrm{d}}

\newcommand{\Lagr}{\mathcal{L}}
\DeclareMathOperator{\tr}{tr}            
\DeclareMathOperator{\IIm}{\mathrm{Im}}  
\newcommand{\hc}{^\dagger}               
\newcommand{\cc}{^\ast}               
\newcommand{\degree}{^\circ}             
\newcommand{\hv}[1]{\hat{\bvec{#1}}}                   
\newcommand{\bvec}[1]{\mathbf{#1}}                     
\newcommand{\gvec}[1]{\boldsymbol{#1}}                 
\newcommand{\twoCol}[2]{\begin{pmatrix} #1 \\ #2 \end{pmatrix}}
\newcommand{\fourCol}[4]{\begin{pmatrix} #1 \\ #2 \\ #3 \\ #4 \end{pmatrix}}
\newcommand{\twoMatrix}[4]{ \begin{pmatrix}{#1} & {#2} \\ {#3} & {#4} \end{pmatrix} }

\DeclareMathOperator{\diag}{\mathrm{diag}}     

\newcommand{\diagPart}{\textrm{diag}}
\newcommand{\offdiagPart}{\textrm{offdiag}}

\begin{document}

\title{Angular instabilities of a homogeneous neutrino gas\\from (pseudo)scalar nonstandard self-interactions}
\author{Oleg~G.~Kharlanov} \email{kharlanov@physics.msu.ru}
\affiliation{Faculty of Physics, Lomonosov Moscow State University, 1/2 Leninskie Gory, 119991 Moscow, Russia}
%


\begin{abstract}
    We demonstrate that in the presence of (pseudo)scalar four-fermion nonstandard interactions, a homogeneous and isotropic gas of Majorana neutrinos
    placed in a strong magnetic field can exhibit an angular flavor instability mixing the two neutrino helicities
    (i.e., a `neutrino-antineutrino' instability). This instability is most pronounced for the inverted mass hierarchy and high neutrino number
    densities; at the same time, even a tiny transition magnetic moment of the order of $10^{-24}\mu_{\text{B}}$ is quite sufficient to trigger it.
    We study the properties of the neutrino-antineutrino mode using the linear stability analysis and then perform a numerical
    simulation to analyze the late-time properties of an oscillating neutrino gas with V--A and (pseudo)scalar interactions. The latter
    analysis reveals three phases of a chaotic, `thermalized' late-time state of neutrinos: along with a low-density phase
    with the SO(3) symmetry broken by V--A interactions and a symmetric high-density phase, another, (pseudo)scalar-induced
    phase with a broken rotational symmetry appears. The phases can be distinguished by the average late-time probabilities
    of (anti)neutrino flavors and/or by the neutrino-to-antineutrino ratio, the latter being a key quantity for the (pseudo)scalar phase.
    We thus show that observations of these quantities from a supernova could place new constraints on the (pseudo)scalar nonstandard
    neutrino interactions.
\end{abstract}

    \maketitle

    \section{Introduction}
    \label{sec:Intro}

    Over the past ten years, experimental breakthroughs have highlighted neutrino oscillations as a
    probe of otherwise inaccessible regions inside the Sun and the Earth itself
    \cite{SK2014_DayNight, SK2016_solarNus, Borexino2011_Be7, Borexino2020_CNO, Borexino2021_review, Salas2019}:
    the Mikheyev--Smirnov--Wolfenstein (MSW) effect leaves observable signatures in the neutrino
    flavor-energy spectra, which carry information on the neutrino creation point and their
    interactions on the way to the detector \cite{MikheevSmirnov1985, Wolfenstein1978}.
    Amongst the latter ones are the so-called nonstandard neutrino interactions beyond the V--A interaction
    from the electroweak sector of the Standard Model~\cite{NSSI_Ohlsson2013, NSSI_Farzan2018, NSSI_Bhupal2019,
    NSSI_Ge2019, NSI_Borexino2020, NSI_IceCube2021, Chatterjee2021_NSIs}.
    Today, the beyond-Standard-Model (BSM) physics manifested in such hypothetical interactions
    can also be tested at a completely new frontier of supernova neutrinos undergoing collective,
    i.e., self-induced flavor oscillations (see, e.g., Refs.~\cite{NSSI_Das2017, NSSI_Dighe2018, NSSI_Yang2018_SPint,
    NSSI_Lei2020, Kharlanov2021}). Indeed, under extreme conditions (neutrino and matter
    densities, superstrong magnetic fields, temperatures, etc.) typical for a supernova explosion,
    the effect of neutrino-neutrino forward scattering results in a strongly nonlinear flavor evolution
    \cite{Mirizzi2016_SNnus, Duan2006_CollOsc, Duan2010_review, Chakraborty2016_review},
    and this evolution, unlike the linear, noncollective one, can be quite sensitive to a number
    of new parameters. Among such parameters are the neutrino magnetic moment \cite{Kharlanov2021, deGouvea2012_2013},
    the nontrivial flavor structure of the V--A interaction \cite{NSSI_Das2017, NSSI_Dighe2018, NSSI_Lei2020},
    the non-V--A four-fermion couplings possibly originating from BSM mediators \cite{NSSI_Yang2018_SPint,
    Kharlanov2021}, as well as the neutrino mass hierarchy and the very nature---Dirac or Majorana---of the neutrino
    \cite{Duan2010_review, deGouvea2012_2013, NSSI_Yang2018_SPint, Kharlanov2021}.

    One of the definitive factors governing collective flavor transformations in supernovae is the dynamical instabilities,
    which are intrinsically present in the corresponding nonlinear evolution equations \cite{Duan2010_review, Raffelt2013_AngularInstabilities,
    Duan2015_NLM, Johns2020_FastInstabilities, Glas2020_FastInstabilities, Capozzi2020_FastInstabilities, Duan2019_DR, Duan2006_CollOsc}.
    On one hand, the flavor instabilities lead to observable signatures in the neutrino spectra, such as the spectral splits
    and the signatures resulting from a chaotic regime of nonlinear flavor evolution \cite{Duan2006_swaps, Duan2019_NLM, Duan2020_NRM, Richers2021_ParticleInCell}.
    On the other hand, the instabilities are by their nature able to amplify small effects, possibly working as
    a `magnifying glass' \cite{Kharlanov2021, deGouvea2012_2013}. In any case, a brute-force numerical treatment of unstable and strongly nonlinear
    integro-differential equations for collective oscillations is virtually impossible in the most realistic setups
    \cite{Duan2010_review, Chakraborty2016_review}. However, it is often instructive to study these oscillations
    within more simplified setups, revealing features that are also expected to hold at least qualitatively in a realistic supernova environment.

    In our recent paper \cite{Kharlanov2021}, we have studied the effects of nonstandard, scalar and pseudoscalar four-fermion neutrino-neutrino
    interactions (further referred to as nonstandard neutrino self-interactions, NSSIs) on the collective flavor evolution of Majorana neutrinos,
    demonstrating that these interactions open a new channel of spin-flavor instabilities. Remarkably, development of such instabilities can be triggered by tiny neutrino
    transition magnetic moments $\mu_{12} \sim 10^{-24}\mu_{\text{B}}$ (i.e., those consistent with the Standard Model \cite{Giunti2009_nuEMP})
    in quite realistic supernova magnetic fields of the order of $B \sim 10^{12}~\text{Gauss}$, and the instabilities themselves
    are fast enough to drastically deform the neutrino spectra within several kilometers above the neutrino sphere. In the Majorana case,
    these instabilities mixing the neutrino states with the two helicities $\alpha = \mp1/2$ (hereinafter referred to as the neutrino and
    antineutrino states, respectively, for brevity)
    should lead to anomalous neutrino-to-antineutrino ratios for the fluxes coming from a supernova and nonstandard spectral splits.
    It should be noted, however, that these results have been obtained within a very simple setup of the so-called single-angle scheme \cite{Duan2010_review},
    which does not consistently account for nontrivial angular neutrino distributions, introducing a certain geometric factor instead \cite{Duan2006_swaps, Dasgupta2008_GeomFactor}.
    Moreover, in fact, a state with a nontrivial mixing of the two neutrino helicities cannot be spherically-symmetric \cite{Kharlanov2021},
    thus, application of the single-angle scheme---a first step beyond the uniform isotropic neutrino gas---to neutrino-antineutrino
    instabilities appears rather qualitative. Obviously, our study~\cite{Kharlanov2021} of the properties of NSSI-induced instabilities should be extended
    to a more general setup allowing for angular degrees of freedom. Such a treatment is given in the present paper.

    As the spherically-symmetric case does not allow for neutrino-antineutrino mixing and this mixing is introduced instead by
    interaction of the neutrino magnetic moment with the external magnetic field
    \cite{deGouvea2012_2013, Dvornikov2012_Heff, Cirigliano2015_spinQK, Kharlanov2019, Abbar2020},
    we focus on a system with one preferred direction in
    space along the magnetic field, namely, on an axisymmetric homogeneous gas of Majorana neutrinos with (pseudo)scalar NSSIs.
    Following quite a standard methodology of studies of collective neutrino oscillations, including our recent work~\cite{Kharlanov2021},
    we start with a linear stability analysis (Sec.~\ref{sec:Stability}) and then complement it with a numerical simulation of
    the flavor evolution (Sec.~\ref{sec:Simulation}), to see what comes beyond the linear regime. The auxiliary
    Sec.~\ref{sec:EvolutionEquation} is devoted to adaptation of the general flavor evolution equations derived in
    Ref.~\cite{Kharlanov2021} to the axisymmetric setup in question. The analysis of instabilities of a homogeneous and isotropic
    neutrino gas presented in Sec.~\ref{sec:Stability} extends the results of Ref.~\cite{Duan2013_Angular} to (pseudo)scalar NSSIs.
    This analysis justifies the presence of an NSSI-induced neutrino-antineutrino instability and
    reveals that its properties qualitatively studied in Ref.~\cite{Kharlanov2021} within the single-angle scheme remain the
    same (up to a factor of order unity) upon inclusion of the angular degrees of freedom. Moreover, a new, previously inaccessible conclusion can be made regarding the angular effects:
    the spherical symmetry of the neutrino gas can be dynamically broken by the magnetic field in the presence of NSSIs, and,
    depending on the neutrino mass hierarchy and the neutrino number density, the NSSI channel of the $\mathrm{SO}(3)$
    symmetry breaking can dominate the Standard-Model one, related to neutral-current V--A interactions.
    The numerical
    simulations, which follow in Sec.~\ref{sec:Simulation}, justify these findings, also demonstrating a characteristic
    transition to chaos via excitation and further `thermalization' of unstable angular modes. The `thermal', late-time
    averages of the neutrino flavor probabilities, that are to be measured by an observer, are strongly affected by the
    flavor chaos and, in the presence of (pseudo)scalar NSSIs, they retain a signature of anomalous neutrino-to-antineutrino ratios.
    This is in line with the effects we observed earlier within the single-angle setup~\cite{Kharlanov2021} and is also akin
    to the chaotic behavior known to occur within non-NSSI setups with translational or rotational degrees of freedom
    \cite{Duan2019_NLM, Duan2020_NRM, Richers2021_ParticleInCell, Mirizzi2015_transInv, Bhattacharyya2020_LateTime, KharlanovGladchenko2021}.
    The implications of this class of effects are then discussed in
    Sec.~\ref{sec:Discussion}; Appendix~\ref{app:LegendreBasis} presents the neutrino evolution equations in the basis of Legendre polynomials
    we use for the numerical analysis.

    \section{Axisymmetric neutrino gas in the presence of NSSIs}
    \label{sec:EvolutionEquation}

    In the present and the following sections, we will study collective oscillations of a  homogeneous Majorana neutrino gas with
    density $n_\nu$ placed in magnetic field $\bvec{B} = B\bvec{e}_z = \const$. For simplicity, as well as to obtain
    several analytical results, we will work in the two-flavor ($N_{\text{f}} = 2$) approximation and
    assume that the neutrinos are monochromatic with energy $E$. In what follows, we will first identify the properties
    of a flavor density matrix $\rho$ of axisymmetric neutrino gas, and then substitute the corresponding Ansatz for $\rho$
    into the equations of motion derived in our recent paper~\cite{Kharlanov2021}, to simplify them in the case studied.
    We use relativistic units $\hbar = c = 1$ throughout the paper, occasionally using an explicit $ct$ notation for the time
    measured in distance units.

    Let us start from a neutrino-neutrino interaction Lagrangian containing NSSIs
    \cite{Giunti_book, NSSI_Yang2018_SPint, Kharlanov2021, Shalgar2019_SecretInteractions, NSSI_Khlopov1988}:
    \begin{equation}\label{L_nunu}
        \mathcal{L}_{\nu\nu} = -\frac{G_{\text{F}}}{\sqrt2}
                             \left\{
                                  : \bigl( \nu_a^\mathrm{T} C \gamma^\mu \frac{1-\gamma_5}{2} \nu_a \bigr)^2 :
                                \; + \; \frac{g_{\text{S}}}{4} :(\nu_a^\mathrm{T} C \nu_a)^2:
                                \; + \; \frac{g_{\text{P}}}{4} :(\nu_a^\mathrm{T} C \gamma_5 \nu_a)^2:
                             \right\},
    \end{equation}
    where summation over the mass index $a = 1, \ldots, N_{\text{f}}$ of Majorana neutrino fields $\nu_a(x)$ is assumed,
    $C = -\ii \gamma^2 \gamma^0$ is the charge conjugation matrix, $\gamma^\mu$ and $\gamma_5$ are the Dirac matrices,
    $G_{\text{F}}$ is the Fermi constant, and colons denote normal ordering of operators.
    The first term in braces describes the neutral-current interaction mediated by a $Z$ boson (see, e.g., Ref.~\cite{Giunti_book}),
    while the second and the third terms represent a scalar and a pseudoscalar NSSIs, respectively, equipped with the corresponding
    dimensionless couplings $g_{\text{S,P}}$.
    As the magnetic field (and the NSSIs) make possible coherent transitions between Majorana neutrino states with different helicities,
    the density matrix $\rho_{f\alpha, f'\beta}(\bvec{p}) = \langle \hat{a}\hc_{f'\beta\bvec{p}} \hat{a}_{f\alpha\bvec{p}} \rangle$
    describing neutrinos with momentum $\bvec{p}$ in our problem contains four nontrivial $N_{\text{f}}\times N_{\text{f}}$ blocks
    that correspond to different helicities $\alpha, \beta = \mp 1/2$ \cite{Kharlanov2021, Cirigliano2015_spinQK},
    \begin{equation}
        \rho(\bvec{p}) \equiv \twoMatrix{\rho_{--}}{\rho_{-+}}{\rho_{+-}}{\rho_{++}}
             \equiv \twoMatrix{\rho_{\nu\nu}}{\rho_{\nu\bar\nu}}{\rho_{\bar\nu\nu}}{\rho_{\bar\nu\bar\nu}},
    \end{equation}
    and matrix elements in each block, in turn, are numbered with flavor indices $f,f' = e, x$. Note that
    the entries of the $\rho$ matrix, specifically, those of its off-diagonal blocks, depend on the complex
    phase included in the definition of the neutrino helicity eigenstates $\chi_\pm(\hv{p})$, where $\hv{p} \equiv \bvec{p} / |\bvec{p}|$
    (for details, see the discussion in Ref.~\cite{Kharlanov2021}).
    Indeed, these two-component column vectors are defined as
    \begin{equation}
        \gvec\sigma\cdot \bvec{p} \chi_\pm(\hv{p}) = \pm |\bvec{p}| \chi_\pm(\hv{p}), \quad
        \chi_\pm\hc(\hv{p}) \chi_\pm(\hv{p}) = 1, \quad \chi_-(\hv{p}) = -\ii\sigma_2 \chi_+\cc(\hv{p}),
    \end{equation}
    where $\gvec\sigma$ denotes the Pauli matrices, so that a single $\hv{p}$-dependent complex phase remains to be fixed.
    The choice of this phase affects the way the neutrino density matrix transforms under rotations and thus, in turn,
    the properties of the density matrix $\rho(\bvec{p})$ expressing the axial symmetry of the neutrino gas.
    We have discussed the transformation law for $\rho$ in Ref.~\cite{Kharlanov2021}, and for a rotation
    about the direction $\bvec{e}_z$ of the magnetic field on $\Delta\varphi$ radians, it reads:
    \begin{equation}\label{rho_transformation}
        \rho(\bvec{p}) \to \rho'(\bvec{p}) =  e^{-\ii \Xi(\hv{p},\Delta\varphi) \mathcal{G}} \rho(R(\bvec{e}_z\Delta\varphi)\bvec{p}) e^{\ii \Xi(\hv{p},\Delta\varphi)
        \mathcal{G}},\qquad \mathcal{G} \equiv \twoMatrix{\idMatrix}{0}{0}{-\idMatrix},
    \end{equation}
    where $R(\bvec{e}_z\Delta\varphi)$ is a $3\times3$ rotation matrix about the $z$ axis and the $\Xi$ phase
    depends on the choice of the helicity eigenstates:
    \begin{equation}
        e^{\ii \sigma_3 \Delta\varphi / 2} \chi_\pm(R(\bvec{e}_z\Delta\varphi)\hv{p}) \equiv
        e^{\pm\ii\Xi(\hv{p},\Delta\varphi)}\chi_\pm(\hv{p}).
    \end{equation}
    As we mentioned in the Introduction, a spherically-symmetric neutrino gas cannot have a nontrivial neutrino-antineutrino mixing, i.e.,
    nontrivial off-diagonal blocks of the density matrix \cite{Kharlanov2021}. In contrast, an axisymmetric gas with
    a single rotation axis $\bvec{e}_z$ allows one to choose the $\chi_\pm$ eigenstates in a way that completely eliminates the $\Xi$
    phase:
    \begin{equation}\label{helicity_eigenstates}
        \chi_+(\hv{p}) = \twoCol{\cos\frac{\vartheta}{2} e^{-\ii\varphi/2}}{\sin\frac{\vartheta}{2} e^{\ii\varphi/2}},
        \quad
        \chi_-(\hv{p}) = \twoCol{-\sin\frac{\vartheta}{2} e^{-\ii\varphi/2}}{\cos\frac{\vartheta}{2} e^{\ii\varphi/2}},
    \end{equation}
    where $\bvec{p} \equiv |\bvec{p}|\;(\sin\vartheta \cos\varphi, \sin\vartheta \sin\varphi, \cos\vartheta)$.
    Now, in the chosen basis with $\Xi = 0$, Eq.~\eqref{rho_transformation} tells us that the density matrix of an axisymmetric
    neutrino gas does not depend on the azimuthal angle $\varphi$. That said, we can relate the density matrix to a function of the
    $\vartheta$ angle:
    \begin{equation}\label{rho_ansatz}
        \rho(\bvec{p}, t) = (2\pi)^3 n_\nu E^{-2}\delta(|\bvec{p}| - E) \varrho(\vartheta, t).
    \end{equation}
    The total number density of neutrinos+anineutrinos $n_\nu = \int \tr\rho(\bvec{p},t) \diff^3p / (2\pi)^3$
    dictates a normalization condition on the `reduced' density matrix $\varrho$:
    \begin{equation}
        \int_0^{\pi} \tr \varrho(\vartheta, t) \cdot 2\pi\sin\vartheta \diff\vartheta = 1,
    \end{equation}
    moreover, further we will assume that the total number of neutrinos+antineutrinos of all flavors is isotropic,
    $\tr \varrho(\vartheta, t) = \const$, so that
    \begin{equation}\label{rho_normalization}
        \tr \varrho(\vartheta, t) = 1/4\pi.
    \end{equation}

    Let us now identify the equations of motion on the neutrino density matrix $\varrho(\vartheta, t)$ describing collective
    oscillations. Majorana neutrino oscillations in magnetic fields in the presence of (pseudo)scalar NSSIs obey evolution equations
    derived in our recent paper \cite{Kharlanov2021}. Both in that and in the present paper, we are mainly interested in the `hidden sector' of
    the instabilities that feature helicity mixing and thus can affect the neutrino-to-antineutrino
    ratios. These instabilities are absent in the Standard Model, being governed by the NSSI coupling $g_+ = (g_{\text{S}} + g_{\text{P}}) / 2$,
    and can be excited, e.g., by an interaction of a small but nonzero neutrino transition magnetic moment with the external magnetic field $\bvec{B}$.
    Despite our primary focus, in the following, we will also keep the other coupling $g_- = (g_{\text{S}} - g_{\text{P}}) / 2$
    and a background with a constant electron density $n_e$ and neutron density $n_n$, for completeness.
    Then the evolution equation on the density matrix $\rho(\bvec{p})$ of a homogeneous neutrino gas reads \cite{Kharlanov2021}:
    \begin{eqnarray}\label{rho_evolution_general}
        \ii\frac{\pd \rho(\bvec{p})}{\pd t} &=& \Bigl[ h_{\text{vac}}(\bvec{p}) + h_{\text{mat}} + h_{\text{AMM}}(\bvec{p}) + h_{\text{self}}(\bvec{p}), \; \rho(\bvec{p}) \Bigr],
        \\
        h_{\text{vac}}(\bvec{p}) &=& \frac{\eta \Delta m^2}{4|\bvec{p}|}\twoMatrix{\mathbb{M}}{0}{0}{\mathbb{M}}, \quad
                    \mathbb{M} \equiv \twoMatrix{-\cos2\theta}{\sin2\theta}{\sin2\theta}{\cos2\theta},
        \\
        h_{\text{mat}} &=& G_{\text{F}} \sqrt2 \; \mathrm{diag}\bigl(n_e - \frac{n_n}{2}, -\frac{n_n}{2}, -n_e + \frac{n_n}{2}, \frac{n_n}{2} \bigr),
        \\
        h_{\text{AMM}}(\bvec{p}) &=& \twoMatrix{0}{\mu_{12} B_\perp(\hv{p})\sigma_2}{\mu_{12}B_\perp\cc(\hv{p})\sigma_2}{0},
        \\
        h_{\text{self}}(\bvec{p}) &=& \frac{G_{\text{F}}\sqrt2}{(2\pi)^3} \int \diff^3p'\;
                    (1 - \hat{\bvec{p}} \cdot \hat{\bvec{p}}')
                    \Bigl\{
                        \tr\bigl(\rho(\bvec{p}')\mathcal{G}\bigr) \mathcal{G}
                        + \bigl( \rho(\bvec{p}') - \rho^{\text{cT}}(\bvec{p}') \bigr)^\diagPart
                    \nonumber\\
                    && \hspace{11em}
                        + g_- \bigl( \rho^{\text{T}}(\bvec{p}') - \rho^{\text{c}}(\bvec{p}') \bigr)^\diagPart
                        + g_+ e^{\ii\Gamma(\hv{p},\hv{p}') \mathcal{G}}\bigl( \rho^{\text{T}}(\bvec{p}') - \rho^{\text{c}}(\bvec{p}') \bigr)^\offdiagPart
                    \Bigr\}.
        \label{h_self_general}
    \end{eqnarray}
    The four $h$ terms in the commutator in Eq.~\eqref{rho_evolution_general} describe the vacuum neutrino oscillations,
    the MSW effect of the background matter, the interaction of the neutrino magnetic moment with the magnetic field,
    and the neutrino-neutrino interaction, respectively. They use the following notations.
    The neutrino mass-squared difference equals $\eta \Delta m^2$, where the sign $\eta = \pm 1$ determines the normal
    or the inverted mass hierarchy (NH or IH), respectively, and $\theta$ is the vacuum mixing angle (in simulations, one typically takes
    $\eta \Delta m^2 = m_3^2 - m_1^2 \approx \pm 2.4 \times 10^{-3}\text{ eV}^2$ and $\theta = \theta_{13} \approx 9\degree$~\cite{PDG}).
    The transition magnetic moment $\mu_{12}$ is the only component that survives for Majorana neutrinos in the two-flavor case \cite{Giunti2009_nuEMP}.
    Moreover, as we have demonstrated in Ref.~\cite{Kharlanov2021}, in the $N_{\text{f}} = 2$ case, the Majorana phase can be
    eliminated from the $g_+$ interaction term in Eq.~\eqref{h_self_general} by a redefinition of phases of the neutrino helicity eigenstates,
    so we ignore it here. Further, the norm of a complex number $B_\perp(\hv{p})$ equals the magnetic field strength perpendicular
    to the neutrino momentum $\bvec{p}$, hence the notation. The phase of this number, as well as the $\Gamma$ phase in the
    collective Hamiltonian~\eqref{h_self_general}, is related to the helicity eigenstates $\chi_\pm(\bvec{p})$ according to
    \begin{gather}
        B_\perp(\hv{p}) \equiv \sqrt2 \gvec\zeta_+(\hv{p})\cdot \bvec{B}, \qquad
        e^{\ii \Gamma(\hv{p},\hv{p}')} (1 - \hv{p} \cdot \hv{p}') \equiv 2\gvec\zeta_+(\hv{p})\cdot \gvec\zeta_+(\hv{p}'),
        \label{Bperp_Gamma_def}\\
        \gvec\zeta_\pm(\hv{p}) \equiv \frac{1}{\sqrt2} \chi\hc_{\mp}(\hv{p}) \gvec\sigma \chi_\pm(\hv{p}).
        \label{zeta_def}
    \end{gather}
    Finally, the collective Hamiltonian~\eqref{h_self_general} features the following operations on matrices:
    \begin{gather}
        \rho^\diagPart \equiv \frac{\rho + \mathcal{G} \rho \mathcal{G}}{2} = \twoMatrix{\rho_{\nu\nu}}{0}{0}{\rho_{\bar\nu\bar\nu}},
        \quad
        \rho^\offdiagPart \equiv \frac{\rho - \mathcal{G} \rho \mathcal{G}}{2} = \twoMatrix{0}{\rho_{\nu \bar\nu}}{\rho_{\bar\nu \nu}}{0},
        \\
        \rho^{\text{c}} \equiv \mathcal{C} \rho \mathcal{C} = \twoMatrix{\rho_{\bar\nu \bar\nu}}{\rho_{\bar\nu \nu}}{\rho_{\nu \bar\nu}}{\rho_{\nu\nu}},
        \quad
        \mathcal{C} \equiv \twoMatrix{0}{\idMatrix}{\idMatrix}{0},
    \end{gather}
    so that charge conjugation $\rho \mapsto \rho^{\text{c}}$ simply swaps the neutrino and antineutrino block lines/columns of
    the density matrix.

    We can now adapt the general equation~\eqref{rho_evolution_general} to our monochromatic and axisymmetric setup~\eqref{rho_ansatz},
    arriving at an equation on the density matrix $\varrho(\vartheta, t)$. First of all, in the helicity
    basis~\eqref{helicity_eigenstates}, Eqs.~\eqref{zeta_def} and \eqref{Bperp_Gamma_def} take the form:
    \begin{gather}
        \gvec\zeta_\pm(\hv{p}) = \frac{1}{\sqrt2}
        \bigl(
            \cos\vartheta \cos\varphi \mp \ii \sin\varphi, \;
            \cos\vartheta \sin\varphi \pm \ii \cos\varphi, \;
            -\sin\vartheta
        \bigr),
        \\
        B_\perp \equiv -B \sin\vartheta, \quad e^{\ii \Gamma(\hv{p},\hv{p}')} (1 - \hv{p} \cdot \hv{p}')
        = C_1(\vartheta,\varphi,\vartheta') \sin\varphi' + C_2(\vartheta,\varphi,\vartheta') \cos\varphi'
        + \sin\vartheta \sin \vartheta',
    \end{gather}
    where $\vartheta', \varphi'$ are the spherical angles corresponding to the momentum vector $\bvec{p}'$,
    and $C_1, C_2$ are certain functions independent of $\varphi'$. Now, the terms containing these functions
    vanish upon integration of the $g_+$ term in the collective Hamiltonian~\eqref{h_self_general} over
    $\diff^3p' = p'^2 \diff{p}'\sin\vartheta'\diff\vartheta'\diff\varphi'$, and only the third term remains.
    Quite similarly, $\hv{p} \cdot \hv{p}' = \cos\vartheta \cos\vartheta' + \sin\vartheta \sin\vartheta' \cos(\varphi - \varphi')$,
    with only the $\cos\vartheta \cos\vartheta'$ term surviving after the $\varphi'$ integration of the V--A and the $g_-$
    contributions to the collective Hamiltonian~\eqref{h_self_general}. The $p'$ integral is lifted by the delta function in the
    Ansatz~\eqref{rho_ansatz}, and we arrive at the desired evolution equation on the reduced density matrix:
    \begin{eqnarray}\label{rho_evolution}
        \ii\frac{\pd \varrho(\vartheta, t)}{\pd t} &=& \bigl[h(\vartheta), \; \varrho(\vartheta)\bigr],
        \quad
        h(\vartheta) = h_{\text{vac}} + h_{\text{mat}} + h_{\text{AMM}}(\vartheta) + h_{\text{self}}(\vartheta),
        \\
        h_{\text{vac}} &=& \frac{\eta\omega}{2} \twoMatrix{\mathbb{M}}{0}{0}{\mathbb{M}}, \quad \omega \equiv \Delta m^2 / 2E,
        \\
        h_{\text{mat}}&=& G_{\text{F}} \sqrt2 \; \mathrm{diag}\bigl(n_e - \frac{n_n}{2}, -\frac{n_n}{2}, -n_e + \frac{n_n}{2}, \frac{n_n}{2} \bigr),
        \\
        h_{\text{AMM}}(\vartheta) &=& -\mathbb{A} \sin\vartheta, \qquad \mathbb{A} \equiv \mu_{12} B \twoMatrix{0}{\sigma_2}{\sigma_2}{0},
        \label{h_AMM}
        \\
        h_{\text{self}}(\vartheta) &=& \mu \int_0^\pi 2\pi\sin\vartheta'\diff\vartheta'
                    \Bigl\{
                        (1 - \cos\vartheta \cos\vartheta') \mathbb{K}(\varrho(\vartheta'))
                        + \sin\vartheta \sin\vartheta' \mathbb{L}(\varrho(\vartheta'))
                    \Bigr\}, \quad \mu \equiv G_{\text{F}}\sqrt2 n_\nu,
        \label{h_self}
        \\
        \mathbb{K}(\varrho) &=& \tr(\varrho\mathcal{G}) \mathcal{G} + \wp^\diagPart + g_- \bigl(\wp^\diagPart\bigr)^{\text{T}}, \quad
        \mathbb{L}(\varrho) = g_+ \bigl(\wp^\offdiagPart\bigr)^{\text{T}}, \qquad \wp \equiv \varrho - \varrho^{\text{cT}}.
    \end{eqnarray}
    Note that we have omitted the time argument of $\varrho$ and of the Hamiltonian terms for brevity.

    Finally, for the sake of the numerical simulations we are making further, it is worth quoting the evolution equations for the
    density matrix in the basis of Legendre polynomials $P_l(\cos\vartheta)$,
    \begin{eqnarray}\label{rho_Legendre_def}
        \varrho(\vartheta, t) = \frac{1}{4\pi}\sum_{l = 0}^\infty \varrho_l(t)P_l(\cos\vartheta).
    \end{eqnarray}
    Note in this regard that isotropy of the total neutrino+antineutrino number distribution (Eq.~\eqref{rho_normalization}) implies
    $\tr \varrho_l(t) = \delta_{l,0}$. The evolution equations take the form (for details, see Appendix~\ref{app:LegendreBasis}):
    \begin{eqnarray}
        \ii\frac{\pd \varrho_l}{\pd t} &=& \bigl[h_{\text{vac}} + h_{\text{mat}}+ \mu \mathbb{K}(\varrho_0), \; \varrho_l \bigr]
                                           - \sum_{l' = 0}^\infty \varsigma_{ll'} \bigl[ \mathbb{A}, \; \varrho_{l'} \bigr]
        - \frac{\mu}{3} \left[ \mathbb{K}(\varrho_1), \; \frac{l}{2l-1} \varrho_{l-1} + \frac{l+1}{2l+3} \varrho_{l+1} \right]
        \nonumber\\
        &+& \mu \left[ \sum_{n=0}^\infty \frac{\varsigma_{n}}{2n+1} \mathbb{L}(\varrho_{n}), \; \sum_{l' = 0}^\infty \varsigma_{ll'} \varrho_{l'} \right],
        \label{rho_evolution_Legendre}
    \end{eqnarray}
    where $\varsigma_l$ and $\varsigma_{ll'}$ are the expansion coefficients of $\sin\vartheta$ and $\sin\vartheta P_{l'}(\cos\vartheta)$,
    respectively, in the Legendre polynomials $P_l(\cos\vartheta)$. As we observe, Eqs.~\eqref{rho_evolution} and~\eqref{rho_evolution_Legendre}
    are systems of differential equations with a quadratic nonlinearity, and development of instabilities arising from this nonlinearity
    is to be studied using linear stability analysis and direct numerical simulation in the two following sections.

    \section{Linear stability analysis}%
    \label{sec:Stability}%

    For the sake of the Lyapunov stability analysis, let us neglect the vacuum mixing angle $\theta$, so that
    $h_{\text{vac}} \to -\frac{\eta \omega}{2} \diag(\sigma_3, \sigma_3)$. Moreover, we set $h_{\text{AMM}}$ to zero
    in the evolution equation~\eqref{rho_evolution}, while the effect of the magnetic field on the unstable flavor evolution will be studied via a perturbed initial condition
    for the density matrix. As a result of these simplifications in the setup, an initially diagonal isotropic neutrino density matrix
    \begin{equation}
        \varrho^{(0)} = \frac{1}{4\pi} \diag(s_{\nu_e}, s_{\nu_x}, s_{\bar\nu_e}, s_{\bar\nu_x}), \quad
        s_{\nu_e} + s_{\nu_x} + s_{\bar\nu_e} + s_{\bar\nu_x} = 1,
    \end{equation}
    does not oscillate, being a stationary solution of the evolution equation~\eqref{rho_evolution}
    both within the electroweak model and in the presence of NSSIs. We assume an
    infinitesimal perturbation $\delta\varrho^{(0)}(\vartheta)$ of the initial density matrix and the flavor evolution $\varrho(\vartheta, t)$
    it generates. The linearized evolution equation on the perturbation $\delta\varrho(\vartheta, t) = \varrho(\vartheta, t) - \varrho^{(0)}$
    takes the form:
    \begin{eqnarray}\label{rho_linearized_nonstationary}
        \ii \frac{\pd \delta\varrho(\vartheta, t)}{\pd t}
        &=& \bigl[ h^{(0)}, \; \delta\varrho(\vartheta,t) \bigr]
        + \bigl[ \delta h(\vartheta,t), \; \varrho^{(0)} \bigr],
        \\
        h^{(0)} &\equiv& h_{\text{vac}} + h_{\text{mat}} + 4\pi \mu \mathbb{K}(\varrho^{(0)}),
        \\
        \delta h(\vartheta, t) &\equiv& \mu \mathbb{K}(\delta\varrho_0(t)) - \frac{\mu \cos\vartheta}{3} \mathbb{K}(\delta\varrho_1(t))
        + \mu \sin\vartheta \int_0^\pi \sin\vartheta' \mathbb{L}(\delta\varrho(\vartheta',t)) \cdot 2\pi\sin\vartheta' \diff\vartheta',
    \end{eqnarray}
    where $\delta\varrho_l(t) = (2l+1)\int_0^\pi \delta\varrho(\vartheta, t) P_l(\cos\vartheta)\cdot 2\pi\sin\vartheta\diff\vartheta$
    are the coefficients of the Legendre expansion of the perturbation $\delta\varrho(\vartheta,t)$, defined analogously to Eq.~\eqref{rho_Legendre_def}
    (see also Appendix~\ref{app:LegendreBasis} quoting the properties of expansions in Legendre polynomials).
    We now proceed with a Lyapunov stability analysis of a stationary spherically-symmetric solution $\varrho^{(0)}(\vartheta, t) = \const$, by searching
    particular solutions of Eq.~\eqref{rho_linearized_nonstationary} of the form $\delta\varrho(\vartheta, t) = \delta\varrho(\vartheta)e^{-\ii\lambda t}$
    and identifying $\kappa \equiv \IIm \lambda$ with their growth rates. One thus arrives at a secular equation
    \begin{gather}\label{rho_linearized}
        \bigl[ h^{(0)}, \; \delta\varrho(\vartheta) \bigr] + \bigl[ \delta h(\vartheta), \; \varrho^{(0)} \bigr] = \lambda \delta\varrho(\vartheta),
        \\
        \delta h(\vartheta) \equiv \mu \mathbb{K}(\delta\varrho_0) - \frac{\mu \cos\vartheta}{3} \mathbb{K}(\delta\varrho_1)
        + \mu \sin\vartheta \int_0^\pi \sin\vartheta' \mathbb{L}(\delta\varrho(\vartheta')) \cdot 2\pi\sin\vartheta'
        \diff\vartheta'.
    \end{gather}

    Note that in the above equation, $h^{(0)}$ and $\varrho^{(0)}$ are block-diagonal matrices. The perturbation
    $\delta\varrho(\vartheta)$ may, of course, contain both diagonal and off-diagonal blocks, but application of $\mathbb{K}, \mathbb{L}$
    and commutation with block-diagonal matrices $h^{(0)}$, $\varrho^{(0)}$ does not mix these blocks. Thus, Eq.~\eqref{rho_linearized} splits into
    two independent secular equations describing block-diagonal and block-off-diagonal, or neutrino-antineutrino,
    perturbations (analogously to the single-angle case studied in Ref.~\cite{Kharlanov2021}):
    \begin{eqnarray}\label{rho_linearized_diag}
        \bigl[ h^{(0)}, \; \delta\varrho^\diagPart(\vartheta) \bigr]
        + \mu \Bigl[\int_0^\pi \bigl( \mathbb{K}(\delta\varrho^\diagPart(\vartheta'))
                                    - \cos\vartheta \cos\vartheta' \mathbb{K}(\delta\varrho^\diagPart(\vartheta'))
                               \bigr)
                    \cdot 2\pi\sin\vartheta'\diff\vartheta',
                    \;
                    \varrho^{(0)}
              \Bigr]
        &=& \lambda \delta\varrho^\diagPart(\vartheta),
        \\
        \bigl[ h^{(0)}, \; \delta\varrho^\offdiagPart(\vartheta) \bigr]
        + \mu \Bigl[ \int_0^\pi \sin\vartheta \sin\vartheta' \mathbb{L}(\delta\varrho^\offdiagPart(\vartheta')) \cdot 2\pi\sin\vartheta'\diff\vartheta',
                    \;
                    \varrho^{(0)}
              \Bigr]
        &=& \lambda \delta\varrho^\offdiagPart(\vartheta),
        \label{rho_linearized_offdiag}
    \end{eqnarray}
    where we have explicitly written $\delta\varrho_{0,1}$ in terms of integral contractions with Legendre polynomials $P_{0,1}$
    to bring forward an integral form of the equation. Indeed, both secular equations take a form analogous to a
    homogeneous Fredholm integral equation of the second kind with a separable kernel, i.e., one of the form
    $K(\vartheta, \vartheta') = \sum_{n=1}^{N} f_n(\vartheta) g_n(\vartheta')$ \cite{IntegralEquations}.
    These equations are solved via an Ansatz, which follows the structure of the kernel:
    \begin{eqnarray}
        \delta\varrho^\diagPart(\vartheta) &=& \frac{1}{4\pi} \delta\varrho_0^\diagPart + \frac{1}{4\pi} \delta\varrho_1^\diagPart \cos\vartheta +
                                              \delta\varrho_\ast^\diagPart(\vartheta),
        \\
        \delta\varrho^\offdiagPart(\vartheta) &=& \frac{1}{4\pi} \delta\varrho_{\sin}^\offdiagPart \sin\vartheta + \delta\varrho_\ast^\offdiagPart(\vartheta),
    \end{eqnarray}
    where $\delta\varrho_\ast^\diagPart(\vartheta)$ is orthogonal to $1$ and $\cos\vartheta$ and
    $\delta\varrho_\ast^\offdiagPart(\vartheta)$ is orthogonal to $\sin\vartheta$ with respect to the scalar product
    $\langle f, g \rangle = \int_0^\pi f(\vartheta) g(\vartheta) \cdot 2\pi \sin\vartheta \diff\vartheta$.
    We now project Eq.~\eqref{rho_linearized_diag} onto $1$, $\cos\vartheta$, and their orthogonal complement and
    Eq.~\eqref{rho_linearized_offdiag} onto $\sin\vartheta$ and its orthogonal complement, arriving at
    \begin{subequations}
        \begin{eqnarray}
            \bigl[ h^{(0)}, \; \delta\varrho^\diagPart_0 \bigr]
            + 4\pi \mu \bigl[\mathbb{K}(\delta\varrho_0^\diagPart), \; \varrho^{(0)} \bigr]
            &=& \lambda \delta\varrho^\diagPart_0,
            \label{secular_rho_diag_0}
            \\
            \bigl[ h^{(0)}, \; \delta\varrho^\diagPart_1 \bigr]
            - \frac{4\pi \mu}{3} \bigl[\mathbb{K}(\delta\varrho_1^\diagPart), \; \varrho^{(0)} \bigr]
            &=& \lambda \delta\varrho^\diagPart_1,
            \label{secular_rho_diag_1}
            \\
            \bigl[ h^{(0)}, \; \delta\varrho^\diagPart_\ast \bigr] &=& \lambda \delta\varrho^\diagPart_\ast,
            \label{secular_rho_diag_ast}
            \\
            \bigl[ h^{(0)}, \; \delta\varrho^\offdiagPart_{\sin} \bigr]
            + \frac{8\pi \mu}{3} \bigl[ \mathbb{L}(\delta\varrho_{\sin}^\offdiagPart), \; \varrho^{(0)} \bigr]
            &=& \lambda \delta\varrho_{\sin}^\offdiagPart,
            \label{secular_rho_offdiag_sin}
            \\
            \bigl[ h^{(0)}, \; \delta\varrho^\offdiagPart_{\ast} \bigr] &=& \lambda \delta\varrho_{\ast}^\offdiagPart.
            \label{secular_rho_offdiag_ast}
        \end{eqnarray}
    \end{subequations}
    Obviously, Eqs.~\eqref{secular_rho_diag_ast}, \eqref{secular_rho_offdiag_ast} have only real eigenvalues, since
    $[h^{(0)}, \; \cdot \; ]$ is a Hermitian map with respect to the Frobenius scalar product of two matrices.
    The three remaining equations \eqref{secular_rho_diag_0}, \eqref{secular_rho_diag_1}, and \eqref{secular_rho_offdiag_sin}
    describe block-diagonal isotropic, block-diagonal dipole, and block-off-diagonal instabilities, which are proportional to
    $1$, $\cos\vartheta$, and $\sin\vartheta$, respectively. The properties of the isotropic unstable modes are well-known
    in the absence of NSSIs, $g_\pm = 0$; they also arise within simplified setups ignoring angular effects, such as the single-angle scheme
    (see, e.g., Refs.~\cite{Duan2007_PendulumAnalysis, Duan2010_review, Duan2006_CollOsc}).
    Dipole instabilities are also well established in the no-NSSI case, in particular, the fact that they typically show up
    in the opposite mass hierarchies with the isotropic instabilities \cite{Duan2013_Angular, Duan2015_SymmetryBreaking},
    due to the opposite signs of the couplings in Eqs.~\eqref{secular_rho_diag_0} and~\eqref{secular_rho_diag_1}. Finally,
    the neutrino-antineutrino instabilities described by Eq.~\eqref{secular_rho_offdiag_sin} are simply absent when NSSIs are switched
    off, since $\mathbb{L} = 0$ in this case. It is also important to note that a transformation $\delta\varrho_{\sin}^\offdiagPart
    \mapsto \mathcal{G} \delta\varrho_{\sin}^\offdiagPart \mathcal{G}$ changes the sign of $g_+$ in
    Eq.~\eqref{secular_rho_offdiag_sin}, thus, the neutrino-antineutrino instability rates are even functions of $g_+$.

    \begin{figure}[h]
        $\begin{array}{ccc}
            \includegraphics[width=5.8cm]{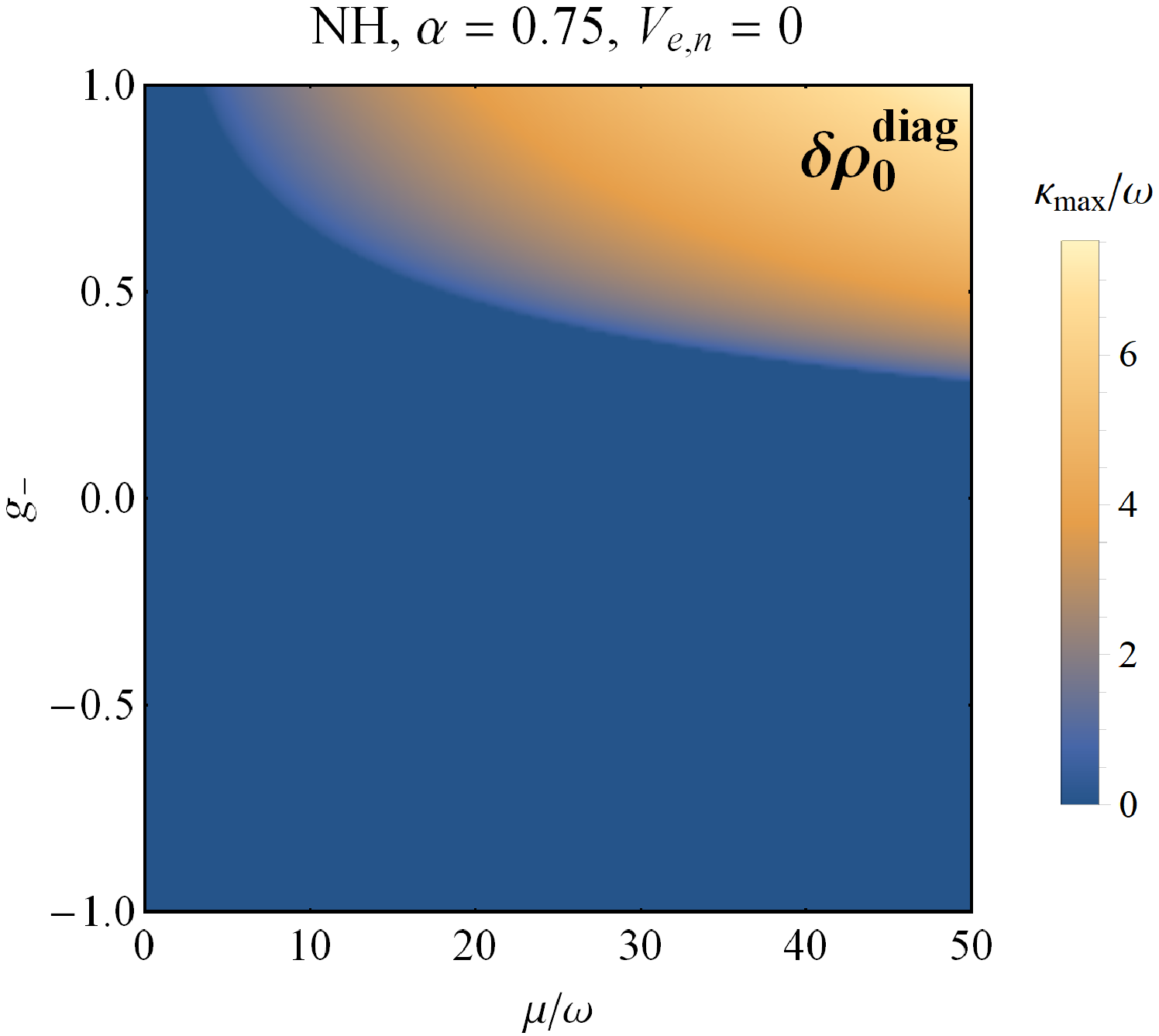} & \includegraphics[width=5.8cm]{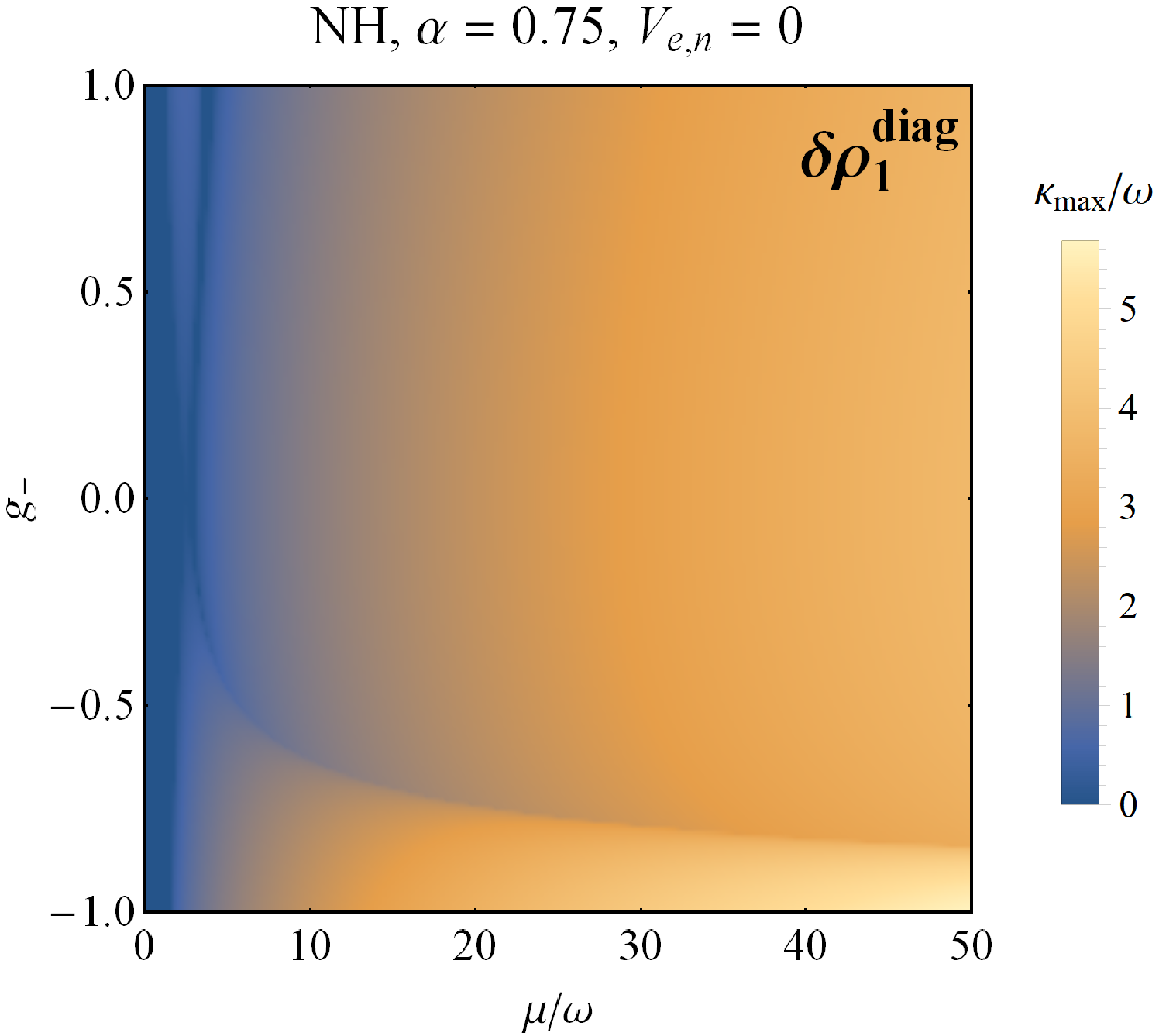} & \includegraphics[width=5.8cm]{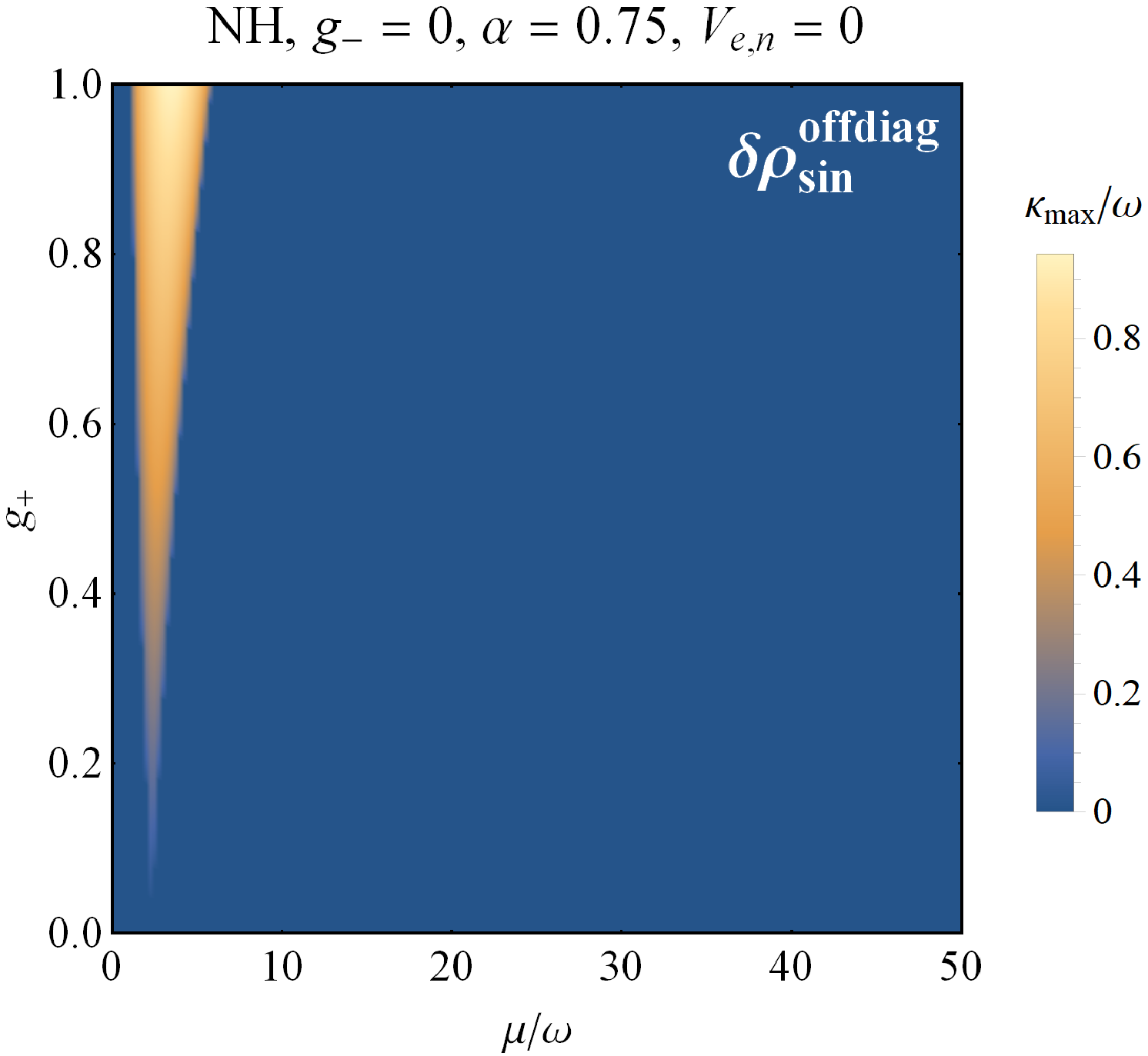} \\
            \text{(a)} & \text{(b)} & \text{(c)} \\[0.5em]
            \includegraphics[width=5.8cm]{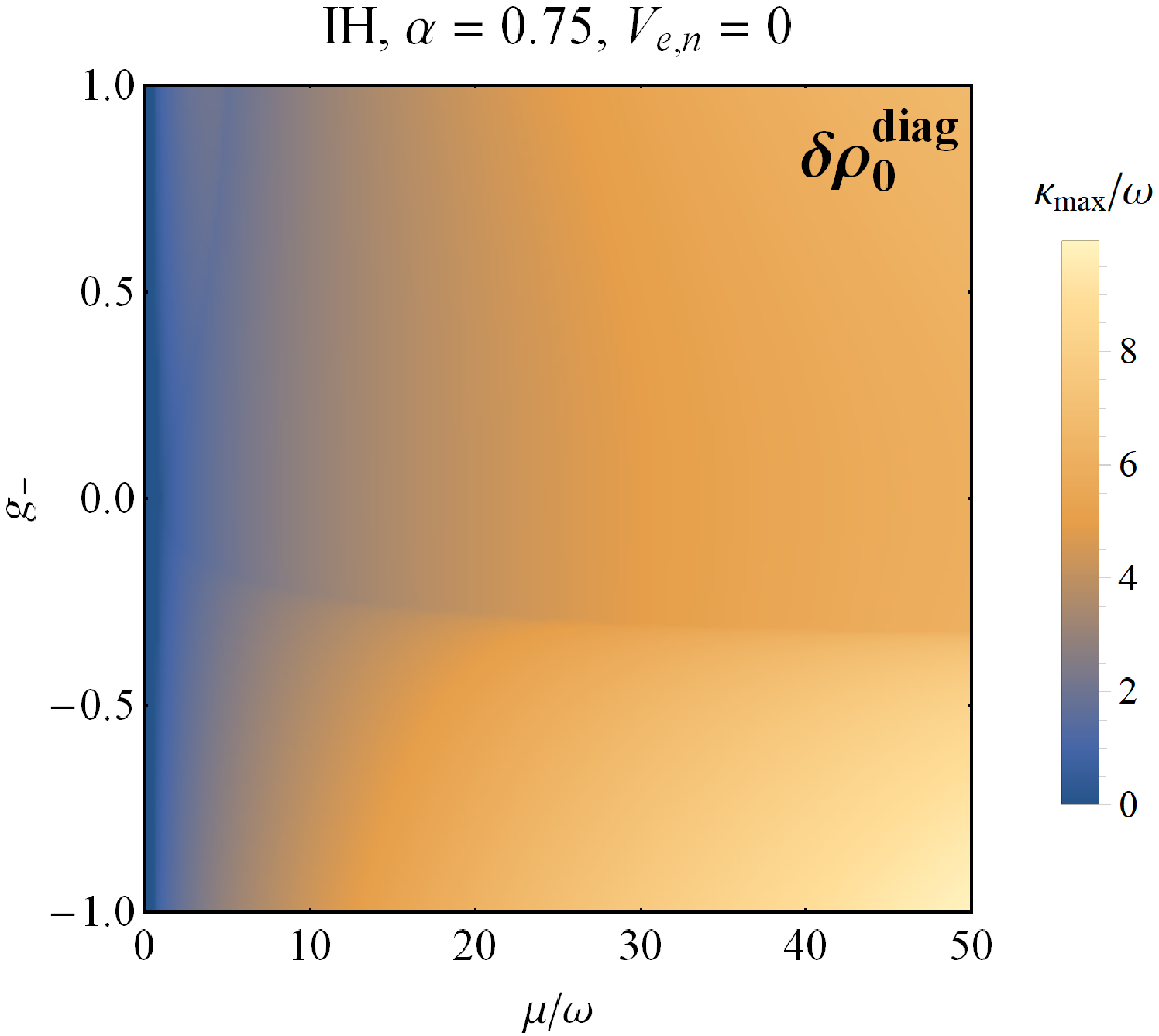} & \includegraphics[width=5.8cm]{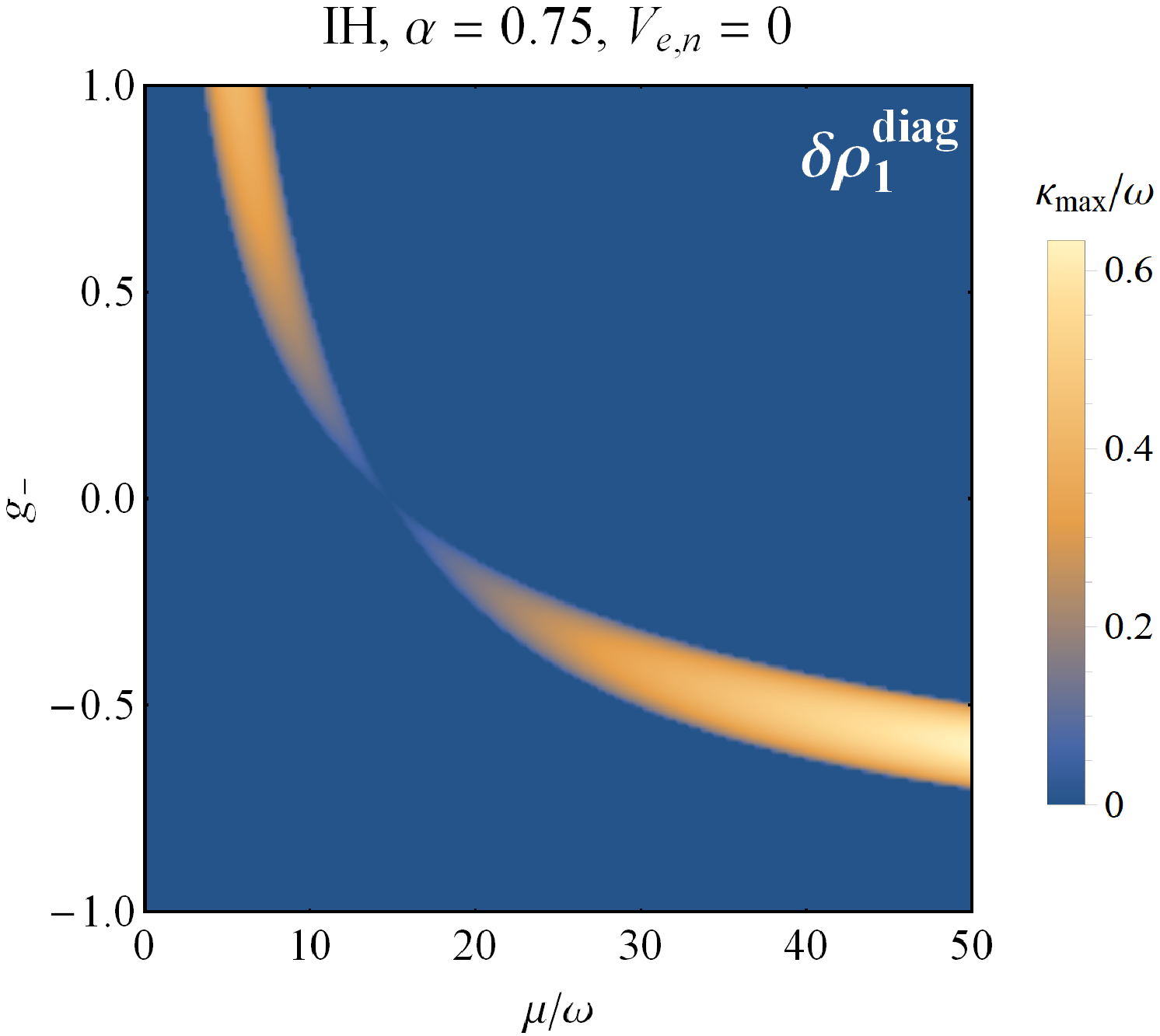} & \includegraphics[width=5.8cm]{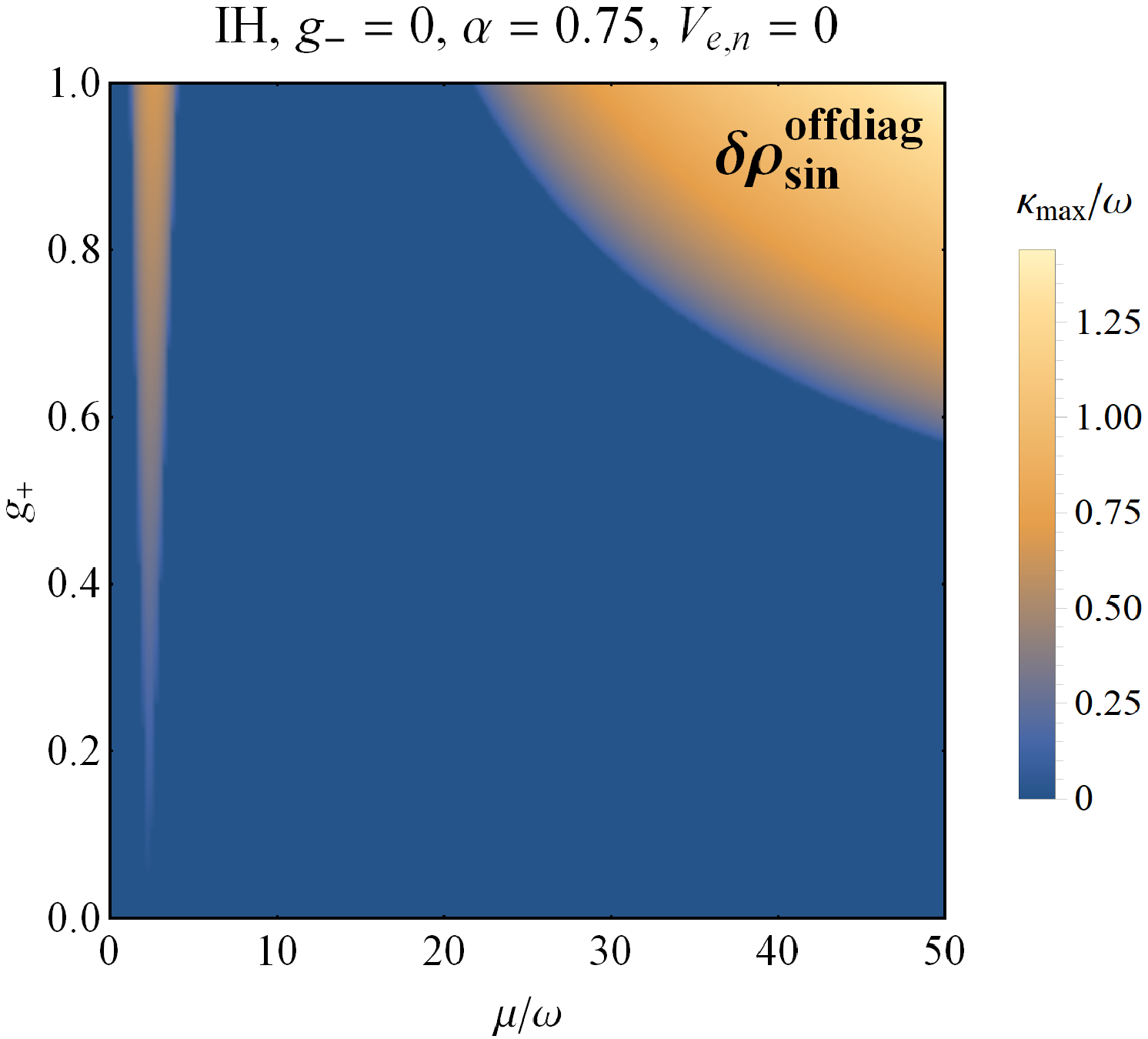} \\
            \text{(d)} & \text{(e)} & \text{(f)}
        \end{array}$
        \caption{Instability growth rates for a monochromatic neutrino flux in the presence of (pseudo)scalar NSSIs for
                (a), (d) isotropic block-diagonal modes, (b), (e) dipole block-diagonal modes, (c), (f) block-off-diagonal
                (neutrino-antineutrino) modes. The upper and the lower row correspond to the normal and inverted hierarchies,
                respectively; the background matter density is set to zero in all panels.}
        \label{fig:instabilityRates_mono}
    \end{figure}

    Now, if (pseudo)scalar NSSIs are present, a nonzero value of the $g_+$ coupling affects only the neutrino-antineutrino
    instabilities~\eqref{secular_rho_offdiag_sin}. In contrast, the $g_-$ coupling affects both the block-diagonal
    modes~\eqref{secular_rho_diag_0}, \eqref{secular_rho_diag_1} and the block-off-diagonal ones~\eqref{secular_rho_offdiag_sin},
    however, its effect on the latter is only through a modification of the diagonal entries of $h^{(0)}$, which does not make a
    qualitative difference from the $g_+ \ne 0, g_- = 0$ case. Moreover, in this case, Eq.~\eqref{secular_rho_offdiag_sin}
    is identical, up to a rescaling of $g_+$ by a factor of $2/3$, to the secular equation we have studied in our recent paper~\cite{Kharlanov2021}
    within a more qualitative reasoning in the spirit of the single-angle scheme. In particular, we have demonstrated there that,
    quite unusually, (a) instabilities of this type can arise even in the absence of antineutrinos and
    (b) they can be matter-enhanced in the presence of the MSW effect, leading to `intermediate' instability rates,
    along with the slow and fast branches \cite{Kharlanov2021}. The latter property contrasts with the Standard Model,
    in which the matter term in the linearized Eq.~\eqref{rho_linearized_nonstationary} can be eliminated by a unitary transformation
    $\delta\varrho(\vartheta, t) = e^{-\ii h_{\text{mat}} t} \delta\tilde\varrho(\vartheta, t) e^{ \ii h_{\text{mat}} t}$ \cite{Kharlanov2021, Duan2010_review}. Regarding the spectrum
    of block-off-diagonal modes, we quote here an analytical expression for the $\lambda$ eigenvalues in the absence of
    background matter ($n_{e,n} = 0$) and for $s_{\nu_e} = 1 / (1+\alpha)$, $s_{\bar\nu_e} = \alpha / (1+\alpha)$,
    $s_{\nu_x} = s_{\bar\nu_x} = 0$, which is a minor modification of a result of Ref.~\cite{Kharlanov2021} allowing for a nonzero $g_-$ coupling:
    \begin{eqnarray}\label{lambda_offdiag}
        \lambda_{1,2,3,4}^{\text{offdiag}} &=& \pm \sqrt{\omega^2 + \left(\frac{1 - \alpha}{1+\alpha}\right)^2 ((3 + g_-)^2 - 2 g_+^2 / 9) \mu^2
                                        \pm \frac{\mu}{1+\alpha} \sqrt{D}
                                        },
        \\
        D &=& \frac{4}{81} g_+^4 \mu^2 \frac{(1-\alpha)^4}{(1+\alpha)^2} + \bigl[4 (3 + g_-)^2 (1-\alpha)^2 + 16 \alpha g_+^2 / 9 \bigr] \omega^2
        + \frac89 (3 + g_-) g_+^2 (1-\alpha)^2 \eta \omega \mu.
        \label{lambda_offdiag_D}
    \end{eqnarray}
    The $\lambda$ eigenvalues for the block-diagonal modes can be found quite analogously.
    Namely, two linear $8\times8$ eigenvalue problems~\eqref{secular_rho_diag_0}, \eqref{secular_rho_diag_1} on the nonzero entries of
    $\delta\varrho^{\text{diag}}_0$ or $\delta\varrho^{\text{diag}}_1$ have four real and four possibly complex eigenvalues.
    The latter ones, determining the stability properties, can be found from a $4\times4$ system:
    \begin{gather}
        \mathcal{M}^{\text{diag}}_l \fourCol{\xi_1}{\xi_2}{\xi_3}{\xi_4} = \lambda \fourCol{\xi_1}{\xi_2}{\xi_3}{\xi_4}, \quad
        \delta\varrho^{\text{diag}}_l \equiv    \begin{pmatrix}
                                                        \xi_5 & \xi_1 &     0 &     0 \\
                                                        \xi_2 & \xi_6 &     0 &     0 \\
                                                        0     &     0 & \xi_7 & \xi_3 \\
                                                        0     &     0 & \xi_4 & \xi_8
                                                    \end{pmatrix},
        \quad l = 0, 1;
        \\
        \mathcal{M}^{\text{diag}}_l =
        \begin{pmatrix}
            \Omega_- - \mu_l \Delta s_{ex} & -\mu_l g_- \Delta s_{ex} & \mu_l g_- \Delta s_{ex} & \mu_l \Delta s_{ex} \\
            \mu_l g_- \Delta s_{ex}  & -\Omega_- + \mu_l \Delta s_{ex} & -\mu_l \Delta s_{ex} & -\mu_l g_- \Delta s_{ex}  \\
            \mu_l g_- \Delta s_{\bar{e}\bar{x}}  & \mu_l \Delta s_{\bar{e}\bar{x}} & -\Omega_+ - \mu_l \Delta s_{\bar{e}\bar{x}} & -\mu_l g_- \Delta s_{\bar{e}\bar{x}} \\
            -\mu_l \Delta s_{\bar{e}\bar{x}} & -\mu_l g_- \Delta s_{\bar{e}\bar{x}} & \mu_l g_- \Delta s_{\bar{e}\bar{x}} & \Omega_+ + \mu_l \Delta s_{\bar{e}\bar{x}}
        \end{pmatrix},
    \end{gather}
    where $\Omega_\pm = \pm \eta \omega + V_e + \mu(1+g_-) (\Delta s_{ex} - \Delta s_{\bar{e}\bar{x}})$, $V_e = G_{\text{F}} \sqrt2 n_e$,
    $\Delta s_{ex} \equiv s_{\nu_e} - s_{\nu_x}$, $\Delta s_{\bar{e}\bar{x}} \equiv s_{\bar\nu_e} - s_{\bar\nu_x}$,  $\mu_0 = \mu$, $\mu_1 = -\mu/3$.
    The analytical expressions for the eigenvalues are more complicated than those for the block-off-diagonal modes [see Eq.~\eqref{lambda_offdiag}],
    so we limit ourselves to quoting only the $\alpha = 1$ case, i.e., the case of equal numbers of electron neutrinos and antineutrinos,
    in which the final expressions get considerably simplified:
    \begin{equation}\label{lambda_diag}
        \lambda_{1,2,3,4}^{\text{diag, }l} = \pm \sqrt{\omega^2 + V_e^2 + \eta \mu_l \omega
                                        \pm \sqrt{4 V_e^2 \omega(\omega + \eta \mu_l) + g_-^2 \mu_l^2 \omega^2}
                                        },
        \quad l = 0, 1.
    \end{equation}
    Within the Standard Model ($g_- = 0$), the above eigenvalues reduce to $\pm\bigl( \sqrt{\omega(\omega + \eta \mu_l)} \pm V_e \bigr)$,
    a result that is known from Ref.~\cite{Duan2013_Angular}. Note that the imaginary parts of the eigenvalues, i.e., the growth rates,
    are insensitive to the matter potential $V_e$ in the $g_- = 0$ case, in accordance with the abovementioned elimination of
    $h_{\text{mat}}$ from Eq.~\eqref{rho_linearized_nonstationary} by a unitary transformation. For a nontrivial $g_-$
    coupling, both the isotropic and the dipole block-diagonal instabilities are affected by the background matter. In this context,
    it is worth mentioning Ref.~\cite{NSSI_Yang2018_SPint}, where the effect of $g_-$ on the collective flavor evolution was studied
    numerically within a neutrino bulb model.

    \begin{figure}[h]
        \includegraphics[width=7cm]{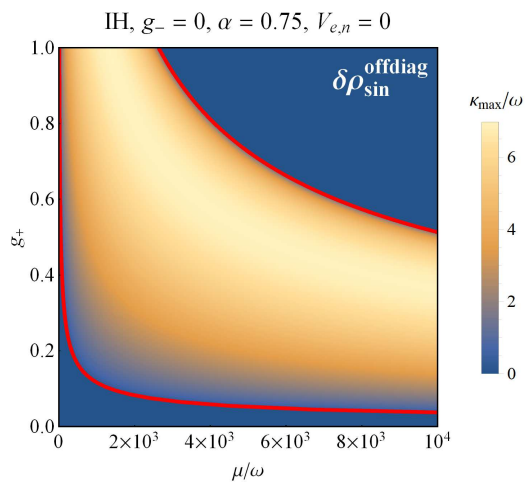}
        \caption{The large-density instability domain for neutrino-antineutrino modes in Fig.~\ref{fig:instabilityRates_mono}(f).
                 The domain exists only for the inverted mass hierarchy. The red boundary lines $\mu = \mu_{\max,\min}(g_+)$ correspond
                 to $D = 0$ in expression~\eqref{lambda_offdiag} for the eigenvalues of the linearized problem.}
        \label{fig:instabilityRates_mono_largemu}
    \end{figure}

    The growth rates of the three types of instabilities are demonstrated in Fig.~\ref{fig:instabilityRates_mono} for the two
    mass hierarchies in the absence of background matter. The neutrino-antineutrino instabilities are shown in
    Fig.~\ref{fig:instabilityRates_mono}(c,~f) for $g_- = 0$ and various $g_+$ values.
    Notably, regardless of the sign of $g_+$, the latter instabilities arise in both hierarchies, whereas the
    isotropic and dipole block-diagonal instabilities [Fig.~\ref{fig:instabilityRates_mono}(a,~d) and Fig.~\ref{fig:instabilityRates_mono}(b,~e),
    respectively] are pronounced in the opposite hierarchies, due to the opposite signs of
    self-couplings in Eqs.~\eqref{secular_rho_diag_0} and~\eqref{secular_rho_diag_1}. In fact, when $g_- = 0$, the isotropic modes
    can become unstable only in the inverted and the dipole modes in the normal hierarchy (though, as we will see in Sec.~\ref{sec:Simulation},
    this statement is exact only in the $\theta = 0$ approximation). One thus observes that
    in the inverted hierarchy, the spherical symmetry of the neutrino gas can be broken by NSSI-induced instabilities.
    The instability domains for the neutrino-antineutrino modes feature a low-density `needle' in both hierarchies and a
    high-density `pool' in the inverted one [see Figs.~\ref{fig:instabilityRates_mono}(c) and~\ref{fig:instabilityRates_mono}(f),
    respectively]. One can reveal that the `pool' corresponds to the densities $\mu$ and the couplings $g_+$ such that the
    discriminant $D < 0$ in Eq.~\eqref{lambda_offdiag}, whereas the `needle' to those parameter values, for which
    $\sqrt{D}$ is real, but the outermost square root in Eq.~\eqref{lambda_offdiag} is taken of a negative number. The
    `needles' extend to arbitrarily small values of $|g_+|$ near a resonance neutrino density
    $\mu_\ast = \frac{\omega (\alpha + 1)}{|\alpha - 1| \cdot |3 + g_-|} \approx 2.33 \omega$ in
    Fig.~\ref{fig:instabilityRates_mono}(c,~f); for this density, the NSSI-induced instability rates are linear in the small
    $|g_+|$ [see Eq.~\eqref{lambda_offdiag}]. The large-density `instability pool' is shown in more detail in
    Fig.~\ref{fig:instabilityRates_mono_largemu}. Its boundary corresponds to $D = 0$, which is a quadratic equation on
    $\mu$ [see Eq.~\eqref{lambda_offdiag_D}]. The two solutions of this equation take an especially simple asymptotic form for $|g_+| \ll 1$:
    \begin{equation}\label{mu_maxmin}
        \mu_{\max,\min}(g_+) \approx \frac{9(3 + g_-)\omega}{g_+^2} \left( \frac{1+\alpha}{1-\alpha} \right)^2
        \frac{(1\pm \sqrt\alpha)^2}{1+\alpha}.
    \end{equation}
    The two curves parameterized by $\mu_{\max,\min}(g_+)$ are also explicitly plotted in Fig.~\ref{fig:instabilityRates_mono_largemu}.
    Importantly, similarly to the `needles', the `pool' extends to arbitrarily small $|g_+|$ couplings for sufficiently large neutrino densities.
    For example, for the parameters chosen in Figs.~\ref{fig:instabilityRates_mono}(f), \ref{fig:instabilityRates_mono_largemu} ($\alpha = 0.75$, $g_- = 0$),
    the minimum neutrino self-coupling to reach the pool is $\mu_{\min} \sim 1.3\times 10^3\omega$ for $g_+ = 0.1$, which
    corresponds to neutrino densities $n_\nu \sim 1.3 \times 10^{30}~\text{cm}^{-3}$. Such densities, in turn, can be reached in a
    supernova explosion with a power $4\pi R_\nu^2 n_\nu E \sim 2\times 10^{50}~\text{erg/sec}$ (the neutrino sphere radius
    chosen is $R_\nu = 50~\text{km}$). One should also note that for such high, but realistic neutrino densities,
    the dipole modes [Fig.~\ref{fig:instabilityRates_mono}(e)] are stable; in fact, the isotropic modes [Fig.~\ref{fig:instabilityRates_mono}(d)]
    also become stable for $\mu/\omega \gtrsim 200$, so the neutrino-antineutrino mode remains the only instability channel.

    \begin{figure}[h]
        $\begin{array}{ccc}
            \includegraphics[width=5.8cm]{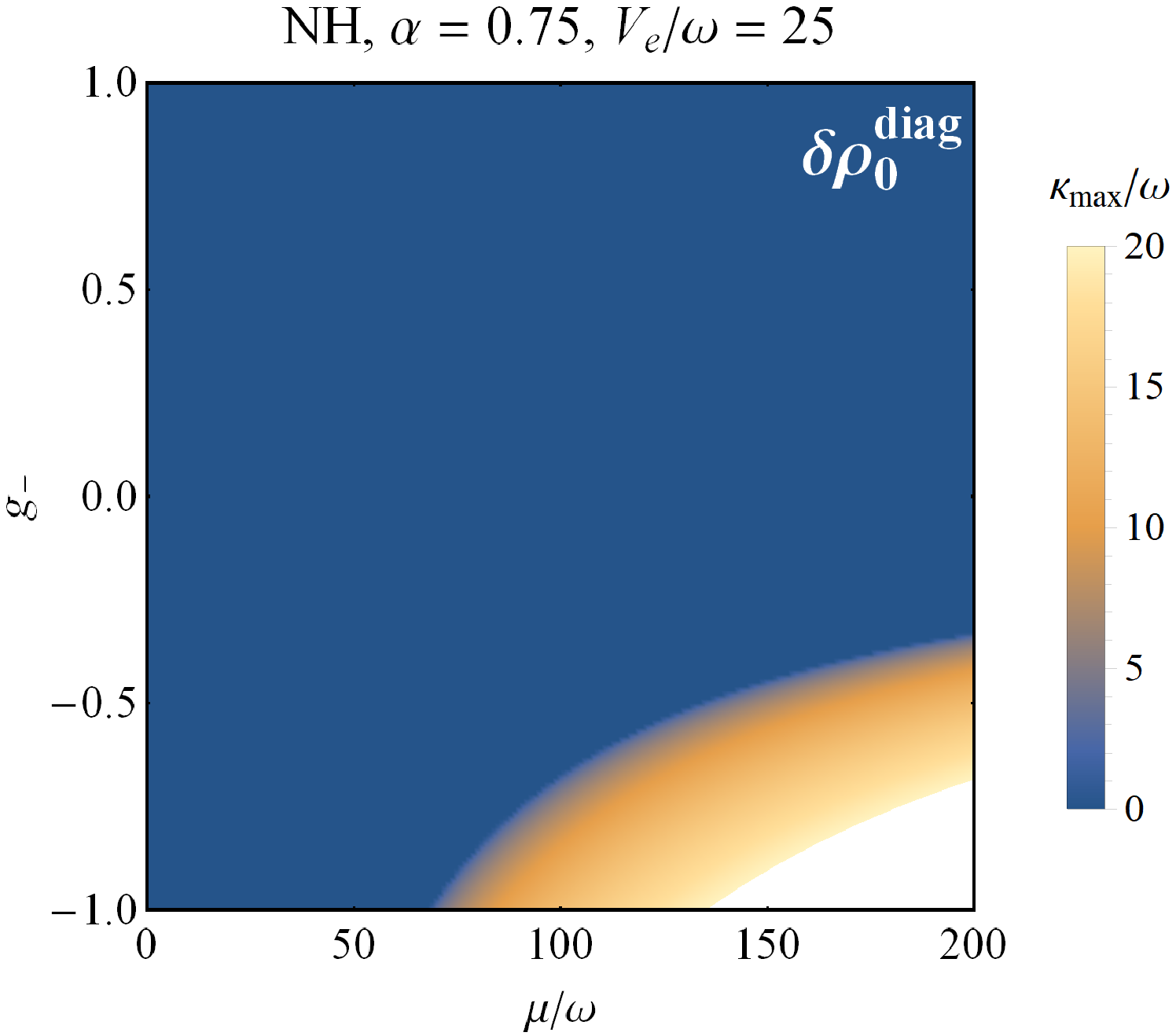} & \includegraphics[width=5.8cm]{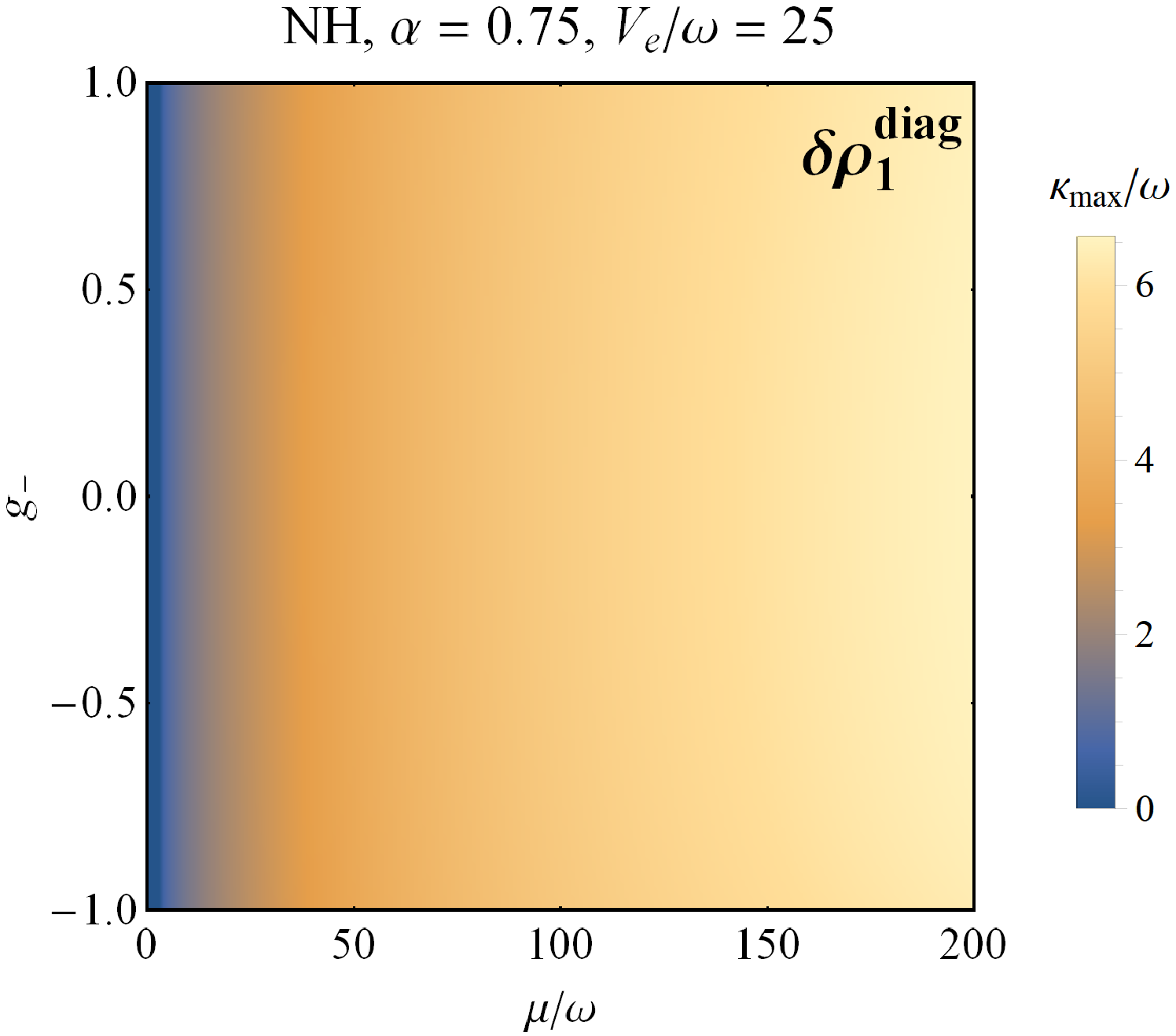} & \includegraphics[width=5.8cm]{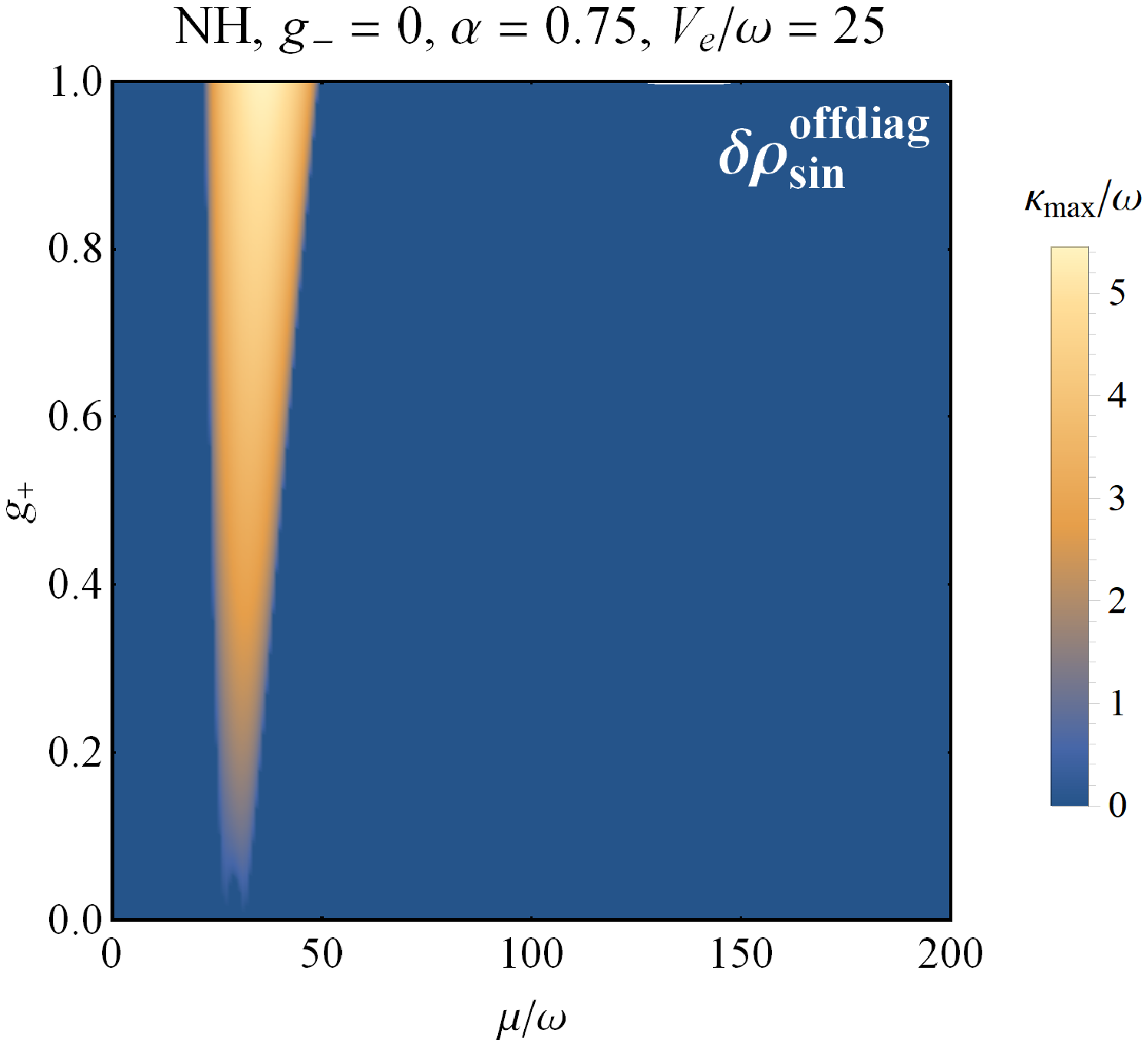} \\
            \text{(a)} & \text{(b)} & \text{(c)} \\[0.5em]
            \includegraphics[width=5.8cm]{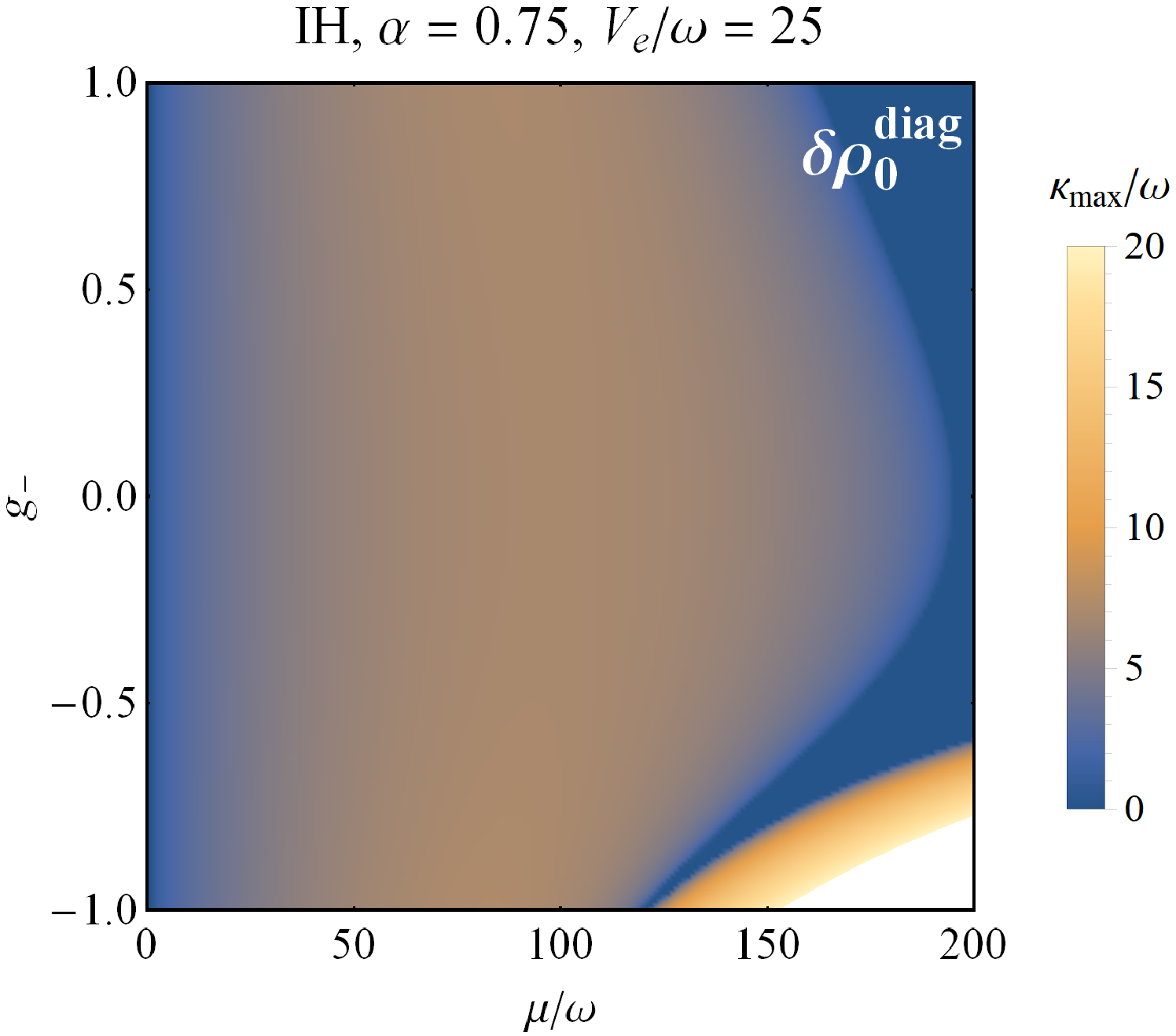} & \includegraphics[width=5.8cm]{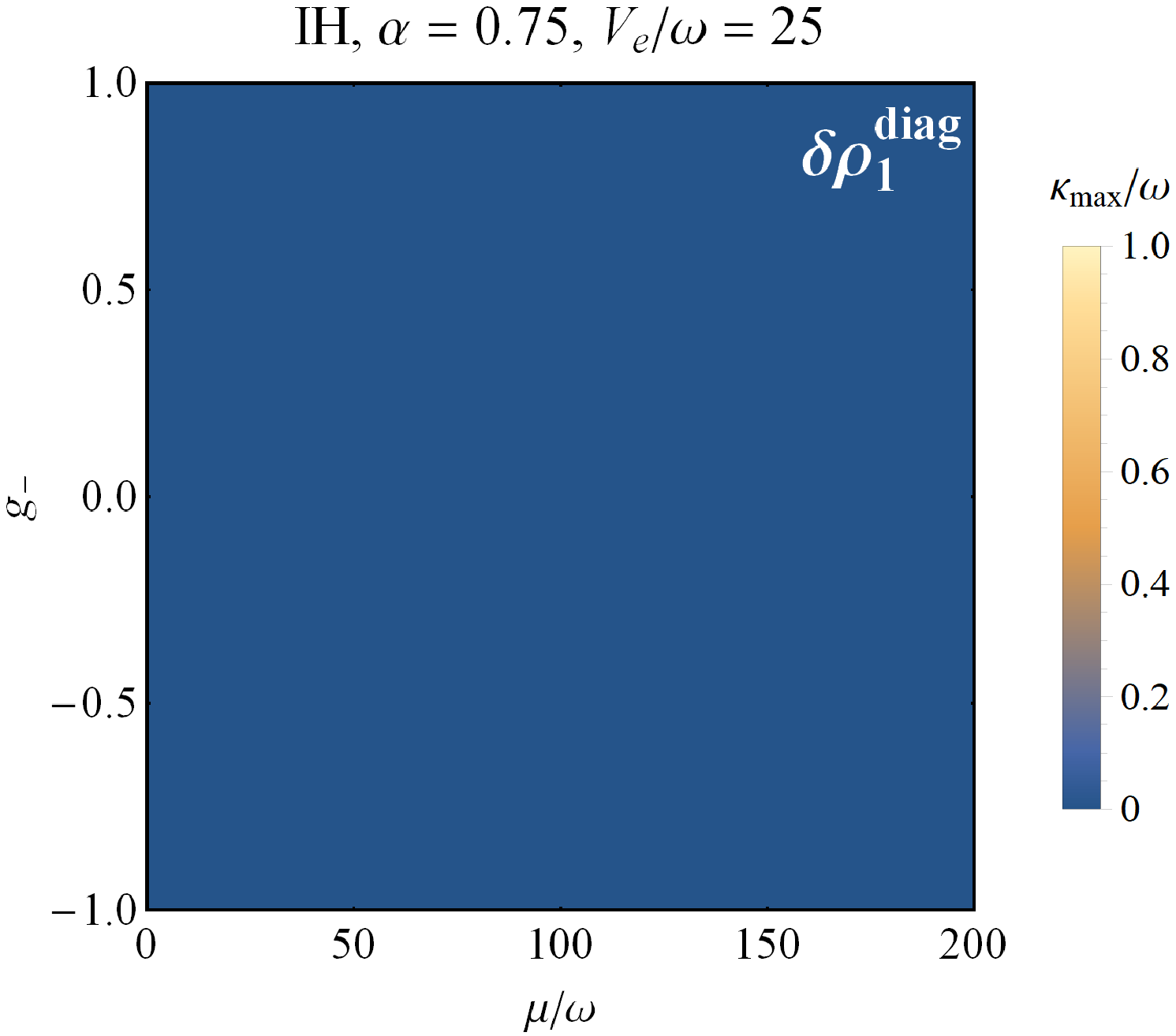} & \includegraphics[width=5.8cm]{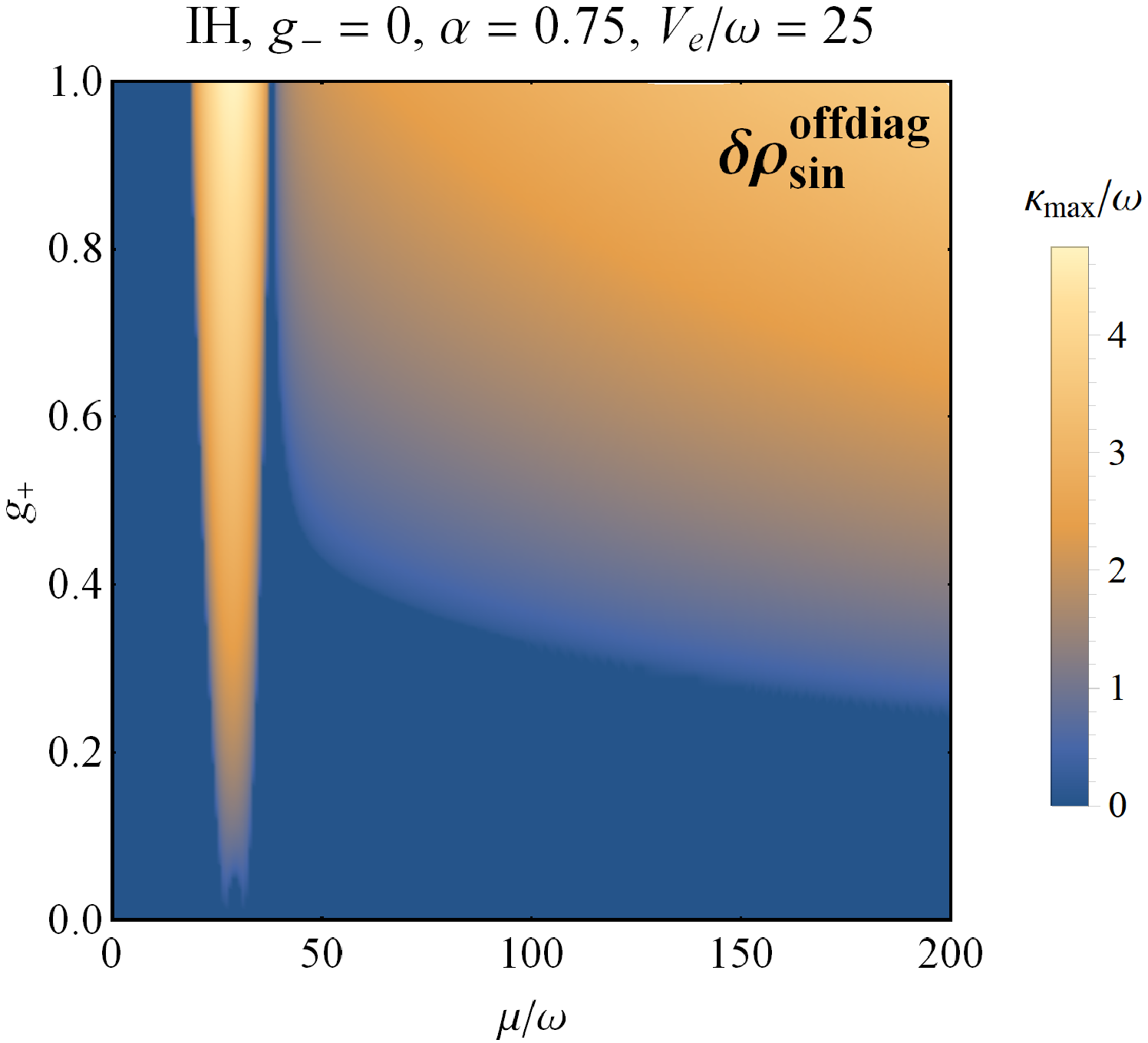} \\
            \text{(d)} & \text{(e)} & \text{(f)}
        \end{array}$
        \caption{The effect of the background matter with $G_{\text{F}} \sqrt{2} n_e = 25\omega$ and $n_n = 1.5 n_e$ on the
                 growth rates of different types of unstable modes. The arrangement of panels (a)--(f)
                 is the same as in Fig.~\ref{fig:instabilityRates_mono}.}
        \label{fig:instabilityRates_mono_matter}
    \end{figure}

    Of course, even a qualitative reference to supernova explosions requires paying attention to the background matter.
    Indeed, Fig.~\ref{fig:instabilityRates_mono_matter} reveals that \emph{in the presence of nontrivial NSSIs}, a background matter potential
    considerably changes the instability rate patterns. Namely, for $G_{\text{F}} \sqrt2 n_e = 25 \omega$ and $n_n = 1.5 n_e$, one
    observes that dipole modes in the inverted hierarchy become stable [strictly speaking, this happens if $\alpha < 1$, e.g., in
    Fig.~\ref{fig:instabilityRates_mono_matter}(e)]. Regarding the neutrino-antineutrino modes, the `instability needles' shift to higher
    neutrino densities, their tips splitting into two, while the `pool' retains its characteristic shape. It is straightforward to see, in
    fact, that for neutrino number densities, such as those shown in Fig.~\ref{fig:instabilityRates_mono_largemu}, the effect of the matter
    potential on the neutrino-neutrino instability is negligible and the rates virtually coincide with those for neutrinos in
    vacuum. Again, in this ultrahigh-neutrino-density regime, both isotropic and dipole block-diagonal modes remain stable,
    and the only flavor instability channel is associated with NSSIs.

    We thus observe that (pseudo)scalar neutrino-neutrino interactions can provide an alternative channel of dynamical
    rotational symmetry breaking of a homogeneous neutrino gas, which can be matter-enhanced and is especially important
    in the case of the inverted mass hierarchy. Moreover, even in the conditions allowing for both the NSSI-induced and the standard,
    V--A instability channels, the NSSI-induced neutrino-antineutrino instability has a unique signature distinguishing
    it from the Standard-Model one: the former one leads to mixing (and further conversion) between neutrinos and antineutrinos, thus breaking the
    SO(3) symmetry of angular distributions of the total (anti)neutrino numbers.

    \section{Numerical simulation}\label{sec:Simulation}

    \begin{figure}[h]
        \includegraphics[width=\textwidth]{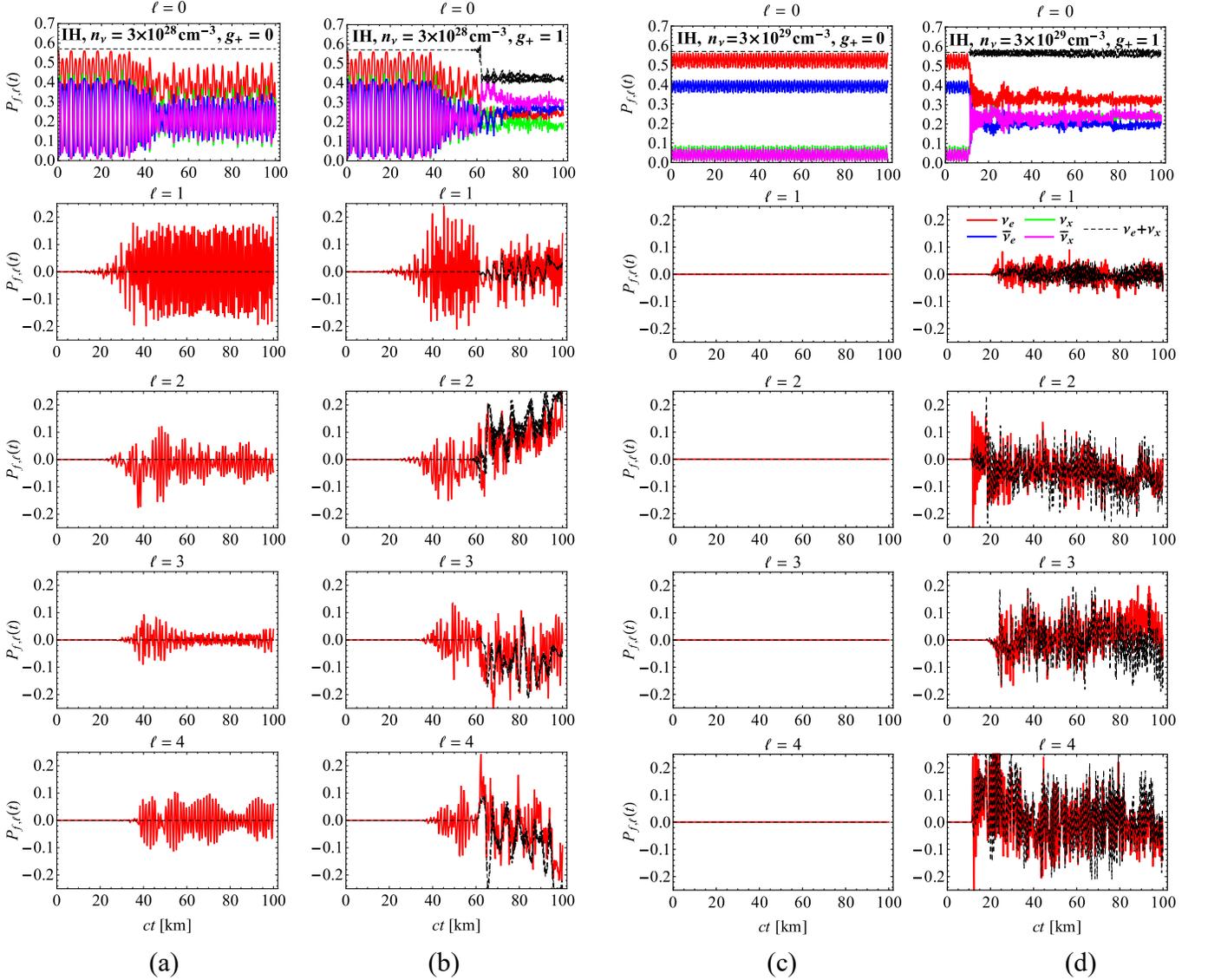}
        \caption{Development of angular instabilities in the neutrino gas with a uniform density $n_\nu$ and the effect of NSSIs on it:
        (a) $n_\nu = 3\times 10^{28}~\text{cm}^{-3}, \; g_+ = 0$, (b) $n_\nu = 3\times 10^{28}~\text{cm}^{-3}, \; g_+ = 1$,
        (c) $n_\nu = 3\times10^{29}~\text{cm}^{-3}, \; g_+ = 0$, (d) $n_\nu = 3\times10^{29}~\text{cm}^{-3}, \; g_+ = 1$.
        The mass hierarchy is inverted in all the panels.
        The plots demonstrate the five lowest harmonics $l = 0,\ldots, 4$ of the (anti)neutrino flavor probabilities, i.e.,
        the diagonal entries of the density matrix components $(\rho_l(t))_{f,f}$; the black dashed line corresponds to
        the harmonics of the total neutrino ($\nu_e + \nu_x$) numbers. In the $l > 0$ plots, only the $\nu_e$ and
        $\nu_e + \nu_x$ curves are shown for vividness.}
        \label{fig:instabilityDevelopment_harmonics}
    \end{figure}

    \begin{figure}[h]
        $$\begin{array}{c}
            \includegraphics[width=\textwidth]{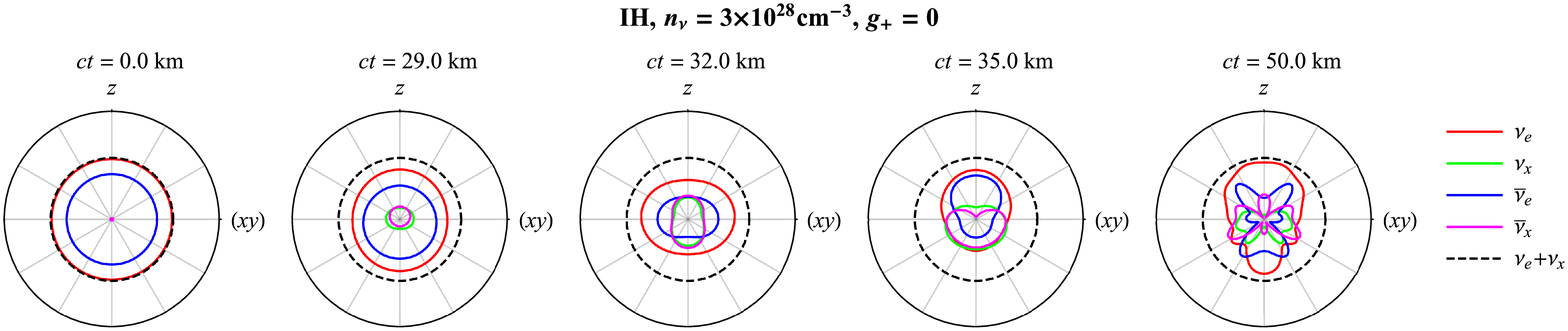} \\
            \text{(a)} \\[1em]
            \includegraphics[width=\textwidth]{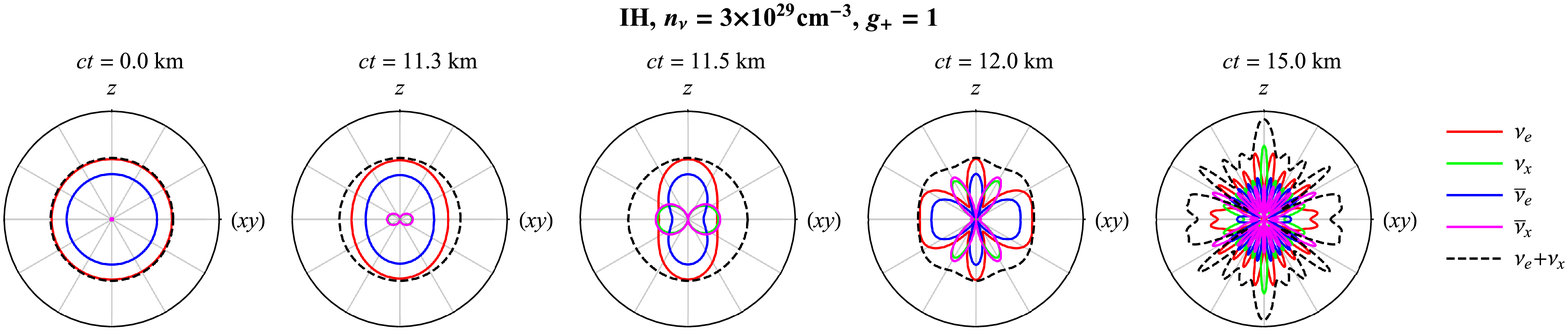} \\
            \text{(b)}
        \end{array}$$%
        \caption{Development of angular instabilities in the neutrino gas with a uniform density $n_\nu$ (inverted hierarchy):
        (a) the dipole block-diagonal instability for $n_\nu = 3 \times 10^{28}~\text{cm}^{-3}$ and $g_+ = 0$, \;
        (b) the $\sin\vartheta$ block-off-diagonal instability for $n_\nu = 3 \times 10^{29}~\text{cm}^{-3}$ and $g_+ = 1$.
        The polar plots show the flavor probabilities $\varrho_{f,f}(\vartheta, t)$ and the notal number of neutrinos
        $\varrho_{e,e}(\vartheta, t) + \varrho_{x,x}(\vartheta, t)$ at selected moments of time $t$.
        Panels (a) and (b) correspond to the evolution of harmonics presented in Fig.~\ref{fig:instabilityDevelopment_harmonics}(a) and
        Fig.~\ref{fig:instabilityDevelopment_harmonics}(d), respectively.}
        \label{fig:instabilityDevelopment_theta}
    \end{figure}

    A simplified stability analysis presented above has revealed a non-isotropic linearly unstable neutrino-antineutrino mode
    $\delta\varrho_{\sin}^\offdiagPart$ breaking spherical symmetry for a nonzero NSSI coupling $g_+$. To complement this analysis,
    we now proceed with a direct numerical simulation of the flavor evolution of a homogeneous neutrino gas, in order to study the fate
    of instabilities beyond the linear regime and in the presence of a nontrivial vacuum mixing $\theta = 9\degree$.
    In our simulations, we fix the neutrino energy to $E = 10~\text{MeV}$, the magnetic field strength to $B = 10^{12}\text{ Gauss} = \const$,
    the transition magnetic moment to $\mu_{12} = 10^{-24}\mu_{\text{B}}$, and ignore the $g_-$ coupling,
    focusing on the effect of the `hidden-sector' NSSI coupling $g_+$. We study the neutrino number densities in the range
    $n_\nu = 10^{28}-10^{30}~\text{cm}^{-3}$, which correspond to $\mu/\omega \sim 10-1000$ and are of the order of the realistic
    neutrino densities for a supernova with explosion power $10^{49}-10^{50}~\text{erg/sec}$ within tens of kilometers above its neutrino sphere \cite{Mirizzi2016_SNnus}.
    The simulations are based on the ODE system~\eqref{rho_evolution_Legendre}, supplied by initial conditions $\varrho_l(0)$
    on the harmonics of the flavor density matrix. In what follows, we will focus on the flavor evolution of an initially almost isotropic neutrino
    gas:
    \begin{equation}\label{rho_initial_generated}
        \varrho_l(0) = \delta_{l,0} \diag(0.56, 0.01, 0.42, 0.01)
                     + \begin{pmatrix}
                           a_l^1 & \frac{a_l^2 + \ii a_l^3}{\sqrt2} & 0 & 0 \\
                           \frac{a_l^2 - \ii a_l^3}{\sqrt2} & a_l^4 & 0 & 0 \\
                           0 & 0 & a_l^5 & \frac{a_l^6 + \ii a_l^7}{\sqrt2} \\
                           0 & 0 & \frac{a_l^6 - \ii a_l^7}{\sqrt2} & -a_l^1 - a_l^4 - a_l^5
                       \end{pmatrix},
        \quad l = 0, 1, \ldots, l_{\max},
    \end{equation}
    where $a_l^1, \ldots, a_l^7 \sim \mathcal{N}\bigl(0, \frac{\epsilon}{2l_{\max}+1} \bigr)$ are a set of independent random
    variables representing the block-diagonal `noise' in the initial condition, which keeps, however, the total number of
    neutrinos and antineutrinos of all flavors isotropic (see Eq.~\eqref{rho_normalization}). The diagonal entries of the reference,
    `noise-free' matrix are chosen to give the $\bar\nu_e$-to-$\nu_e$ ratio $\alpha = 0.42 / 0.56 = 0.75$, i.e., the value used within
    the linear stability analysis in the previous section, while the remaining, $\nu_x$ and $\bar\nu_x$ entries are set to small but
    nonzero values, $0.01$, to keep the eigenvalues of $\varrho(\vartheta, 0)$ nonnegative after adding the noise.
    In our simulations, we use the noise amplitude $\epsilon = 10^{-3}$ for $l_{\max} + 1 = 26$ harmonics.

    In the following subsections, we discuss different features of the numerical solutions obtained, in particular,
    in light of the predictions of the linear stability analysis.

    \subsection{Development of angular NSSI-induced instabilities}
    \label{sec:Simulation_A}%

    \begin{figure}[h]
        \begin{center}
            \includegraphics[width=8.6cm]{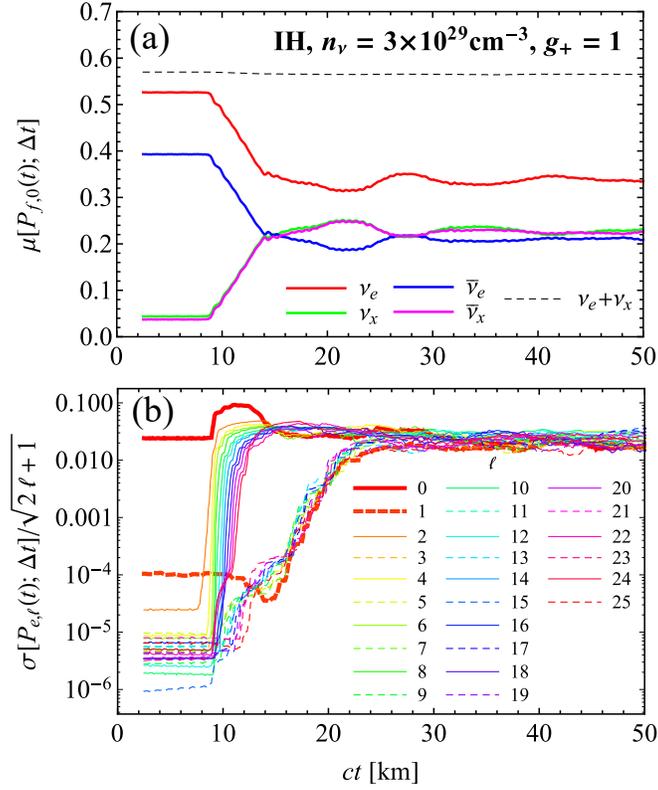}
        \end{center}
        \caption{(a) Moving averages of the isotropic ($l = 0$) harmonic of the four (anti)neutrino probabilities and of the total
        $\nu_e + \nu_x$ probability; (b) moving standard deviations of all the harmonics $l$ of the $\nu_e$ probability, divided by $\sqrt{2l+1}$.
        The averages/standard deviations at the moment of time $t$ are calculated from a segment $[t-\Delta t/2, t + \Delta t/2]$
        with $c \Delta t = 5\text{ km}$. The parameters $n_\nu = 3 \times 10^{29}~\text{cm}^{-3}$, $g_+ = 1$, IH match those in
        Figs.~\ref{fig:instabilityDevelopment_harmonics}(d),~\ref{fig:instabilityDevelopment_theta}(b).}
        \label{fig:thermalization}
    \end{figure}

    \begin{figure}[h]
        \begin{gather*}
            \begin{array}{ccc}
                \includegraphics[height=5cm]{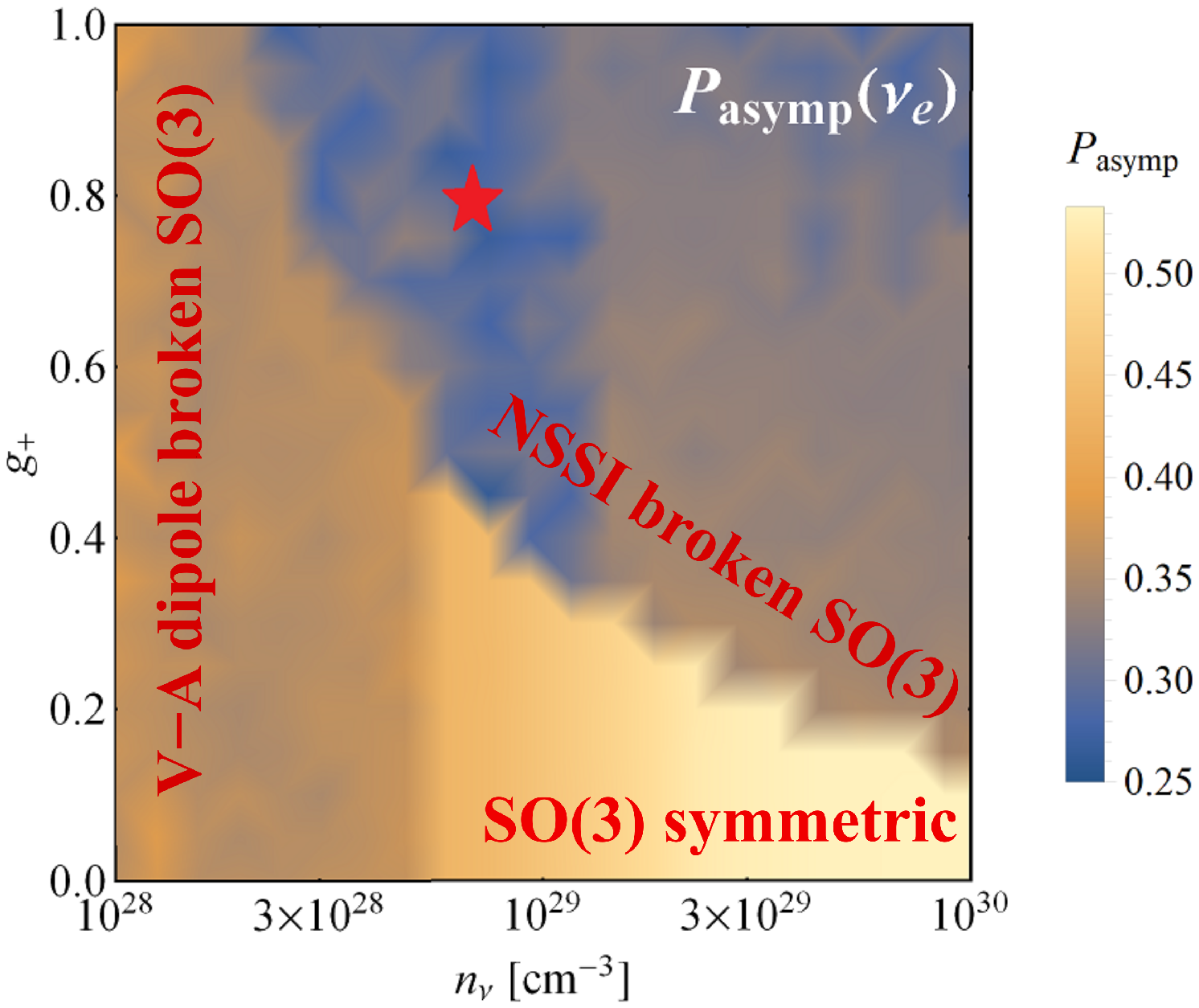} & \includegraphics[height=5cm]{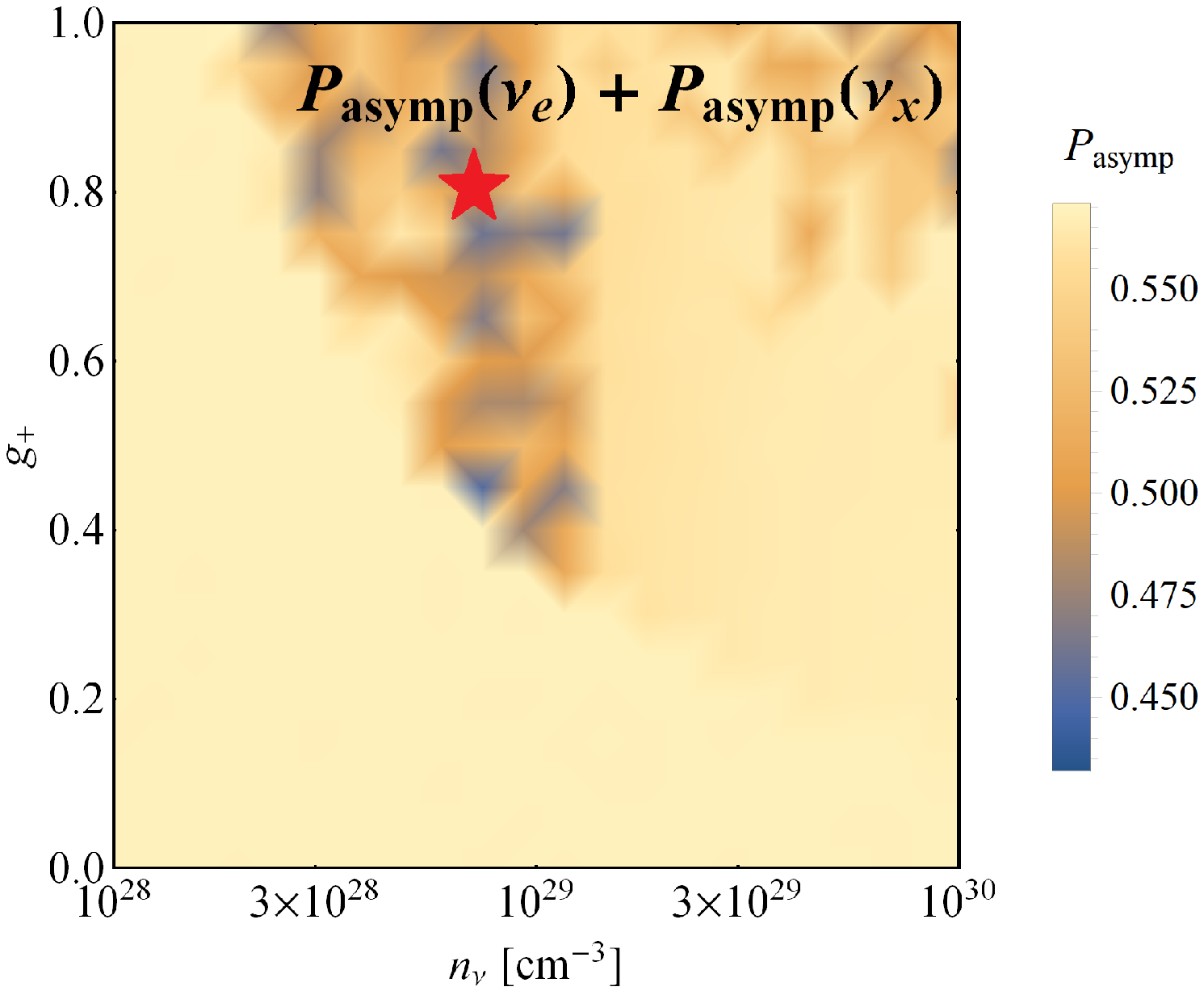} & \includegraphics[height=5cm]{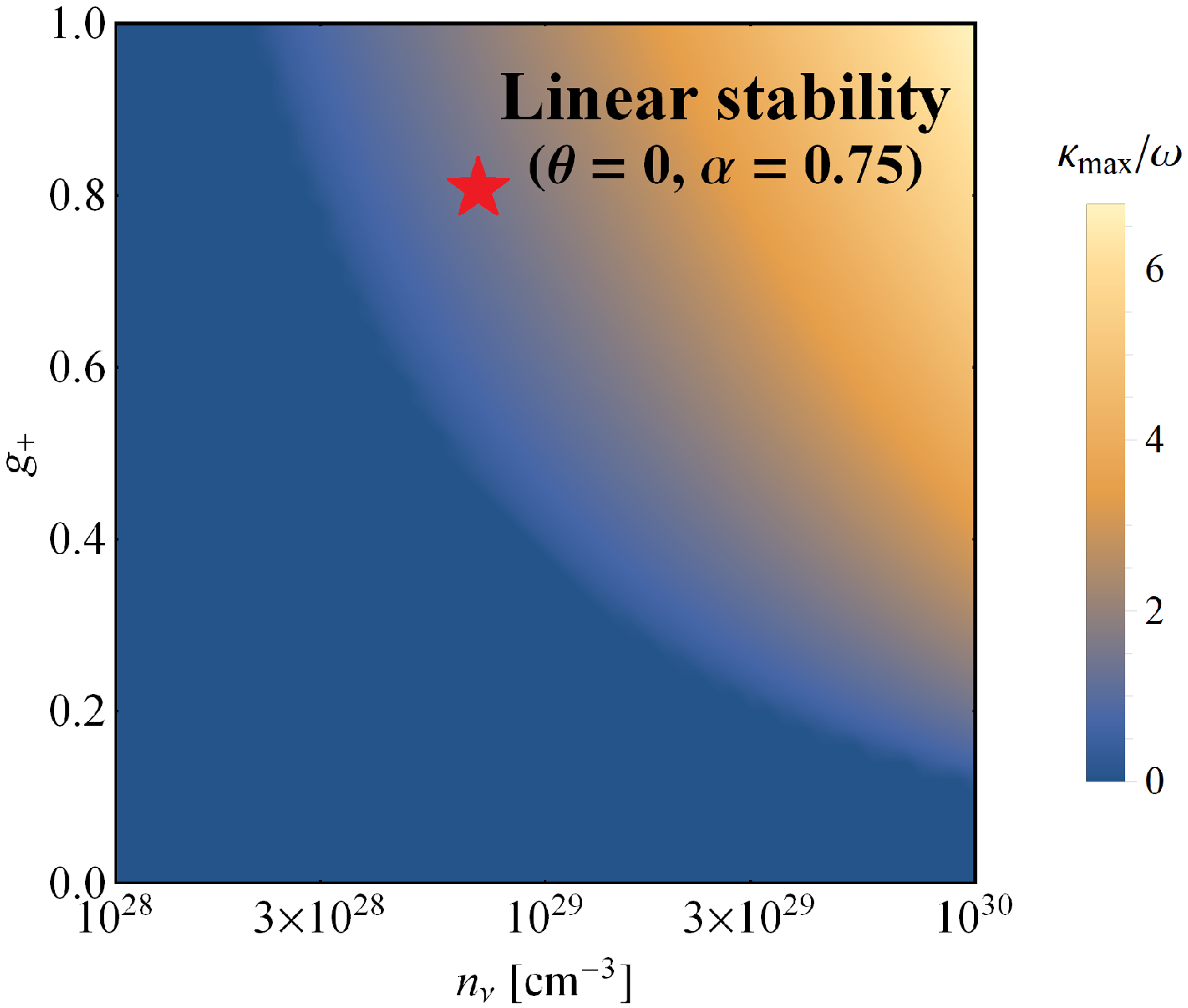}\\
                \text{(a)} & \text{(b)} & \text{(c)}
            \end{array}
            \\
            \begin{array}{cc}
                \includegraphics[height=5cm]{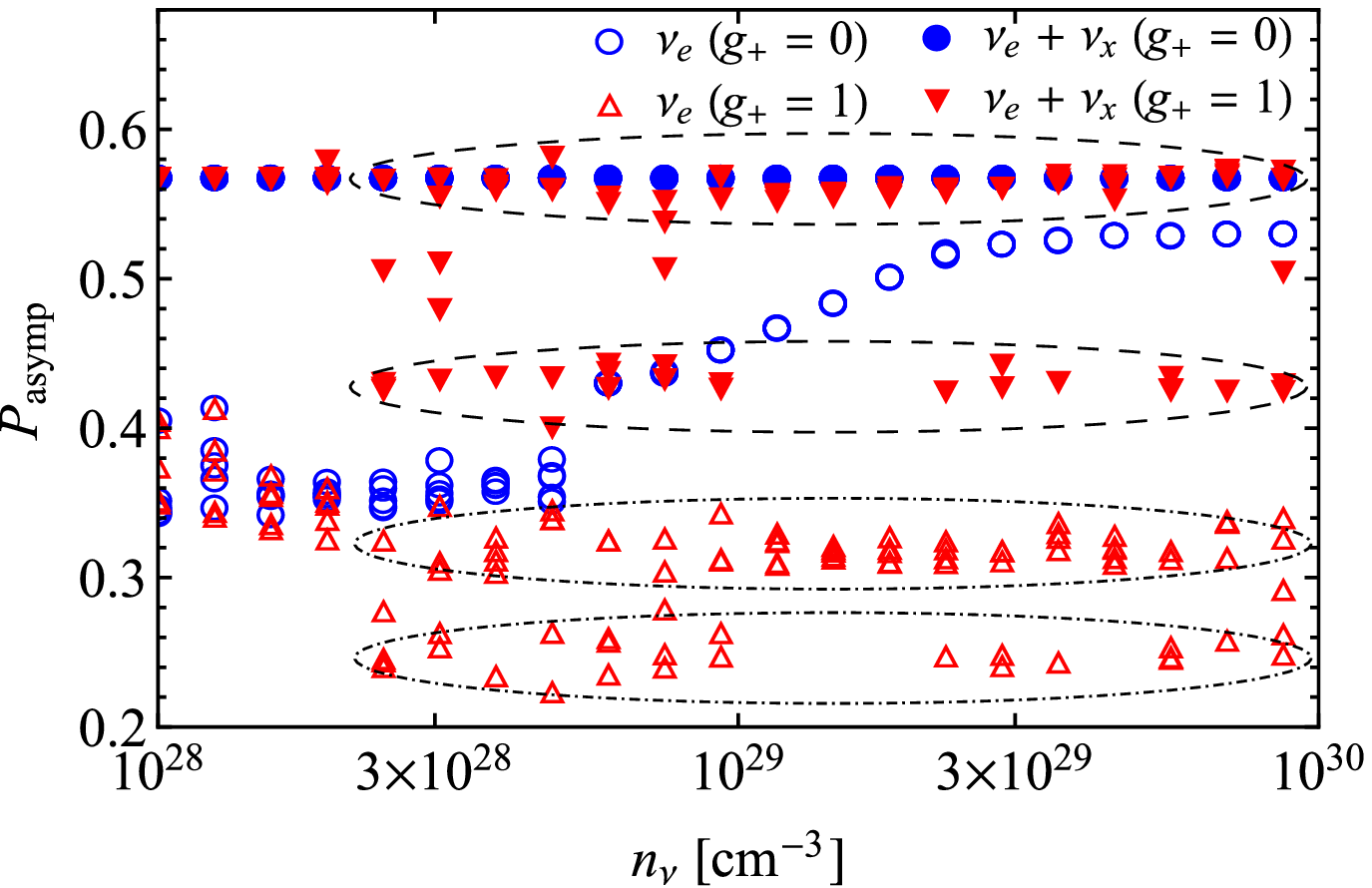} & \includegraphics[height=5cm]{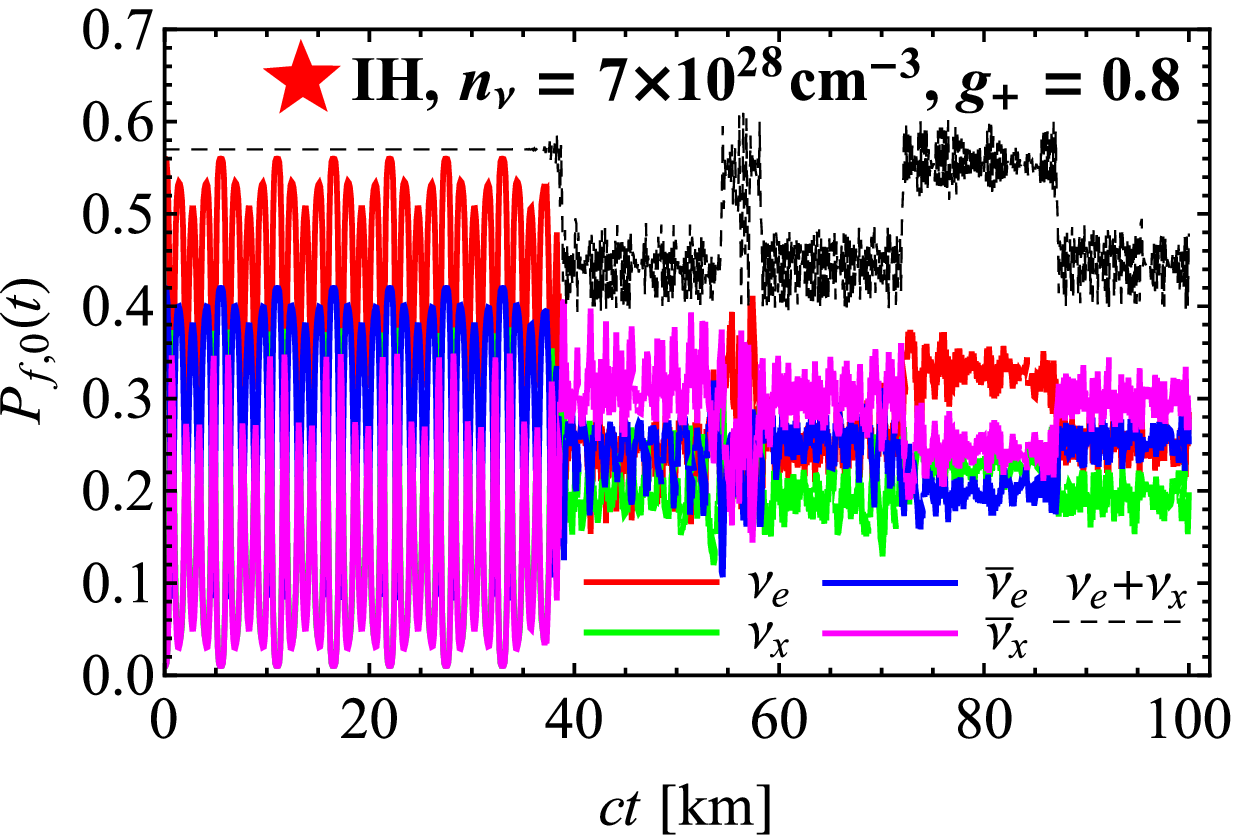}\\
                \text{(d)} & \text{(e)}
            \end{array}
        \end{gather*}%
        \caption{Average late-time neutrino flavor probabilities for a homogeneous neutrino gas with different
        densities $n_\nu$ and NSSI couplings $g_+$ (inverted mass hierarchy). Panels (a), (b) show the probabilities of $\nu_e$ and the two neutrino flavors $\nu_e +
        \nu_x$ together, respectively. For a comparison, panel (c) demonstrates the instability rates for the block-off-diagonal mode
        from the linear stability analysis [Figs.~\ref{fig:instabilityRates_mono}(f),~\ref{fig:instabilityRates_mono_largemu}].
        Every probability point $P_{\text{asymp}}(n_\nu, g_+)$
        in panels (a), (b) results from averaging the neutrino probability $P_{f,0}(t)$ over a late-time interval $ct \in [90~\text{km}, 100~\text{km}]$
        and a further averaging over 5 ODE solutions with randomly generated noise~\eqref{rho_initial_generated} in the initial
        density matrix. The results of individual runs are presented in panel (d) for $g_+ = 0 \text{ and } 1$, revealing pairs of
        quasi-stable $\nu_e$ and $\nu_e + \nu_x$ probability regions for $g_+ = 1$ (highlighted by two pairs of dashed ellipses). A typical solution
        demonstrating a bistable behavior is plotted in panel (e), with the corresponding neutrino gas parameters marked with a star
        in panels (a)--(c).}
        \label{fig:Pasymp}
    \end{figure}

    A typical picture of SO(3) symmetry breaking by angular instabilities is presented in Fig.~\ref{fig:instabilityDevelopment_harmonics}
    in the presence of NSSIs ($g_+ = 1$) and in the Standard Model ($g_+ = 0$), for different neutrino densities.
    The figure demonstrates the five lowest Legendre harmonics of the flavor probabilities, namely, the harmonics of the diagonal entries
    of the density matrix, $P_{f,l}(t) = \bigl(\varrho_l(t)\bigr)_{f,f}$, as well as those of the total probabilities of the
    two neutrino flavors $\nu_e + \nu_x$. We have focused on the inverted mass hierarchy, since it is expected from
    Sec.~\ref{sec:Stability} that isotropy-breaking instabilities should be suppressed in it in the absence of NSSIs
    [see Figs.~\ref{fig:instabilityRates_mono}(e), \ref{fig:instabilityRates_mono_matter}(e)].
    Regarding the neutrino number densities, we have chosen two representative values $n_\nu = 3 \times 10^{28}~\text{cm}^{-3}$
    and $3 \times 10^{29}~\text{cm}^{-3}$ in Fig.~\ref{fig:instabilityDevelopment_harmonics}(a, b) and (c, d), respectively,
    since they illustrate the two scenarios of isotropy violation: the one due to V--A interactions and the
    high-density scenario of SO(3) breaking driven by (pseudo)scalar NSSIs.

    Although the linear analysis has predicted no dipole instabilities in the Standard Model [$g_\pm = 0$,
    see Fig.~\ref{fig:instabilityRates_mono}(e)] in the case of inverted mass hierarchy,
    Fig.~\ref{fig:instabilityDevelopment_harmonics}(a) clearly reveals that this is no more the case once a nonzero vacuum mixing angle comes into play.
    However, for higher neutrino densities, the anisotropic $l \ne 0$ modes do remain stable, indeed, in the absence of NSSIs
    [Fig.~\ref{fig:instabilityDevelopment_harmonics}(c)].
    Now, when (pseudo)scalar NSSIs are turned on, the picture changes dramatically: for lower densities, the
    neutrino-antineutrino instability develops on top of the dipole instability [Fig.~\ref{fig:instabilityDevelopment_harmonics}(b)],
    while in the high-density regime, the entire development of the $l \ne 0$ harmonics owes itself to this new instability channel
    [Fig.~\ref{fig:instabilityDevelopment_harmonics}(d)]. Specifically, in the latter case, the neutrino-antineutrino $\sin\vartheta$ instability
    leads to an explosive excitation of the even harmonics, which further propagates to the odd-$l$ modes via secondary instabilities.
    The isotropic, zeroth harmonic also undergoes a dramatic change at the moment of the first `splash' of the anisotropic harmonics:
    quasiperiodic, synchronous oscillations give way to a chaotic, noisy time evolution of flavor probabilities,
    and the very mean values of these noisy dependencies switch to new positions. Notably, for moderate neutrino densities,
    such as those shown in Fig.~\ref{fig:instabilityDevelopment_harmonics}(b), the total number of neutrinos (the dashed line),
    initially being conserved to a high accuracy, abruptly crashes down to a new plateau---exactly at the moment $ct\sim 60~\text{km}$
    when the neutrino-antineutrino mixing reaches an order-of-unity magnitude as a result of exponential growth of NSSI-induced
    block-off-diagonal instabilities. The development of the characteristic dipole ($\cos\vartheta$) and $\sin\vartheta$ instability
    shapes is also evident from Fig.~\ref{fig:instabilityDevelopment_theta}(a) and Fig.~\ref{fig:instabilityDevelopment_theta}(b),
    respectively, which present the angular $\varrho_{f,f}(\vartheta, t)$ dependencies rather than their harmonics.
    Note that the neutrino-antineutrino instability in Fig.~\ref{fig:instabilityDevelopment_theta}(b) develops very quickly,
    within a kilometer, and soon leads to a chaotic pattern. For completeness, let us also mention that in the case of the normal hierarchy,
    the flavor evolution is dominated by the dipole block-diagonal instability for both $g_+ = 0$ and $g_+ = 1$ and no sizeable
    neutrino-antineutrino conversion takes place for the chosen neutrino densities [cf.~the corresponding instability patterns
    from the linear stability analysis in Fig.~\ref{fig:instabilityRates_mono}(b, c)].

    \subsection{Thermalization and equipartition}
    \label{sec:Simulation_B}%

    After the initial stage of their development, the angular instabilities `thermalize', forming a chaotic steady state similar to a
    fully developed hydrodynamic turbulence. The stages of this process are especially vivid for the higher density $n_\nu = 3 \times
    10^{29}~\text{cm}^{-3}$ and $g_+ = 1$: Fig.~\ref{fig:thermalization} shows the moving averages and standard deviations
    for different harmonics $P_{f,l}$ of the neutrino probabilities.
    Namely, Fig.~\ref{fig:thermalization} is obtained from Fig.~\ref{fig:instabilityDevelopment_harmonics}(d) by
    taking the mean values $\mu_f(t)$ of the isotropic harmonic $P_{f,0}$ and the standard deviations $\sigma_l(t) /\sqrt{2l+1}$ of
    all $\nu_e$ harmonics $P_{e,l}$ on different segments $[t - \Delta t / 2, t + \Delta t / 2]$ of width $c \Delta t = 5\text{ km}$.
    The reason for rescaling of $\sigma_l(t)$ by $\sqrt{2l+1}$ was that approximate equality of different $\sigma_l$'s would
    imply equipartition of the corresponding variances $\sigma_l^2$ among the degrees of freedom on the sphere ($2l+1$ per each
    angular momentum $l$). Now, the stages of the thermalization in Fig.~\ref{fig:thermalization} are the following.
    First, in the linear regime, the NSSI-induced instability remains block-off-diagonal and thus does not affect the
    probabilities (i.e., the diagonal entries of the density matrix), which corresponds to the horizontal segments of
    the curves in Fig.~\ref{fig:thermalization}(a,~b). Next, around $ct = 10\text{ km}$,
    the evolution of the diagonal blocks gets drastically modified by the exponentially growing even harmonics, originating from the $\sin\vartheta$
    instability and corresponding to straight, almost vertical lines in Fig.~\ref{fig:thermalization}(b). The even harmonics
    quickly saturate, and the excitation propagates to the odd ones [dashed lines in Fig.~\ref{fig:thermalization}(b)],
    which start a slower but also exponential growth. Finally, around $ct = 30\text{ km}$, an equilibration is
    achieved, which exhibits (at least, approximately) the properties of a steady state with equipartition between different
    spherical harmonics. The latter phenomenon is quite general and has been observed in collective neutrino oscillations within a number
    of setups and using different methods \cite{Duan2019_NLM, Duan2020_NRM, Abbar2020, Richers2021_ParticleInCell, Bhattacharyya2020_LateTime,
    KharlanovGladchenko2021, Sasaki2021}.
    For us, it is important that formation of the steady state is characterized by the new plateaus of the neutrino flavor probabilities in
    Fig.~\ref{fig:thermalization}(a) [and, for lower neutrino densities, by the new values of the $\nu/\bar\nu$ ratio,
    see Fig.~\ref{fig:instabilityDevelopment_harmonics}(b)].

    It is also instructive to study how the properties of the asymptotic, late-time states depend on the parameters $\mu$, $g_+$ of the
    neutrino gas, in other words, to study the equilibrium equation of state for this self-interacting gas. Fig.~\ref{fig:Pasymp}
    demonstrates the results of a numerical simulation: for each $(n_\nu, g_+)$ point, we have randomly generated the
    initial perturbations~\eqref{rho_initial_generated} for 5 runs, and averaged the resulting probabilities $P_{f,0}(t)$
    over a late-time segment $ct \in [90~\text{km}, 100~\text{km}]$. The resulting `phase diagram' is most vivid for the
    electronic neutrino probabilities $P_{e,0}$, and one observes that in the presence of NSSIs and sufficiently high densities,
    the SO(3)-symmetric phase gives way to a phase with a broken spherical symmetry [and different asymptotic $\nu_e$ probabilities,
    see Fig.~\ref{fig:Pasymp}(a)]. Not surprisingly, the transition between the two phases occurs quite precisely at the line predicted by the linear stability
    analysis [Fig.~\ref{fig:Pasymp}(c)], even though the latter was done in the $\theta = 0$ approximation. For lower neutrino gas
    densities, a competition takes place between the block-diagonal dipole channel of symmetry breaking caused by V--A interactions and
    the block-off-diagonal $\sin\vartheta$ channel opened by (pseudo)scalar interactions. Here, the phase boundary also neatly
    agrees with the linear stability analysis (note, however, that the latter, being restricted to the $\theta = 0$ case, could
    not predict the dipole instabilities in the inverted mass hierarchy); the NSSI-dominated phase features a pronounced deficit
    of electron neutrinos, as well as an overall conversion of neutrinos into antineutrinos [see the blue areas in
    Fig.~\ref{fig:Pasymp}(a) and the corresponding domains in Fig.~\ref{fig:Pasymp}(b)]. Interestingly, these blue areas,
    which lie near the boundary between the V--A- and the NSSI-dominated phases, exhibit a behavior resembling hydrodynamic
    intermittency: the fully developed chaotic flavor evolution is switching between two plateaus of the total neutrino ($\nu_e +
    \nu_x$) probability [see Fig.~\ref{fig:Pasymp}(d,~e)]. As a result, a neutrino gas with such $n_\nu, g_+$ parameters can be found in
    different quasi-steady states near $ct = 100~\text{km}$ for slightly different initial density
    matrices~\eqref{rho_initial_generated}, as evidenced by Fig.~\ref{fig:Pasymp}(d).
    Interestingly, Fig.~\ref{fig:Pasymp}(e) also reveals that switching events in such a bistable chaotic system are
    characterized not only by transitions between the two plateaus of the total $\nu_e + \nu_x$ probability,
    but also by an individual  $\nu_{e,x} \leftrightarrow \bar\nu_{x,e}$ probability exchange.
    Such an exchange results in a reflection of the total neutrino probability
    $P_{e,0} + P_{x,0} \mapsto P_{\bar{x},0} + P_{\bar{e},0} = 1 - (P_{e,0} + P_{x,0})$, which is indeed observed in Fig.~\ref{fig:Pasymp}(e).
    In principle, this suggests that the bistability could be intrinsically related to an exchange symmetry
    $\varrho \mapsto \begin{pmatrix} 0 & \sigma_1 \\ \sigma_1 & 0 \end{pmatrix} \varrho \begin{pmatrix} 0 & \sigma_1 \\ \sigma_1 & 0 \end{pmatrix}$,
    which is indeed respected by the equations of motion for $\omega \to 0$ but is broken by the vacuum term. However,
    this interesting issue is beyond the scope of the present paper, so we limit ourselves to pointing out our observation.

    \begin{figure}[h]
        \begin{center}
            \includegraphics[width=\textwidth]{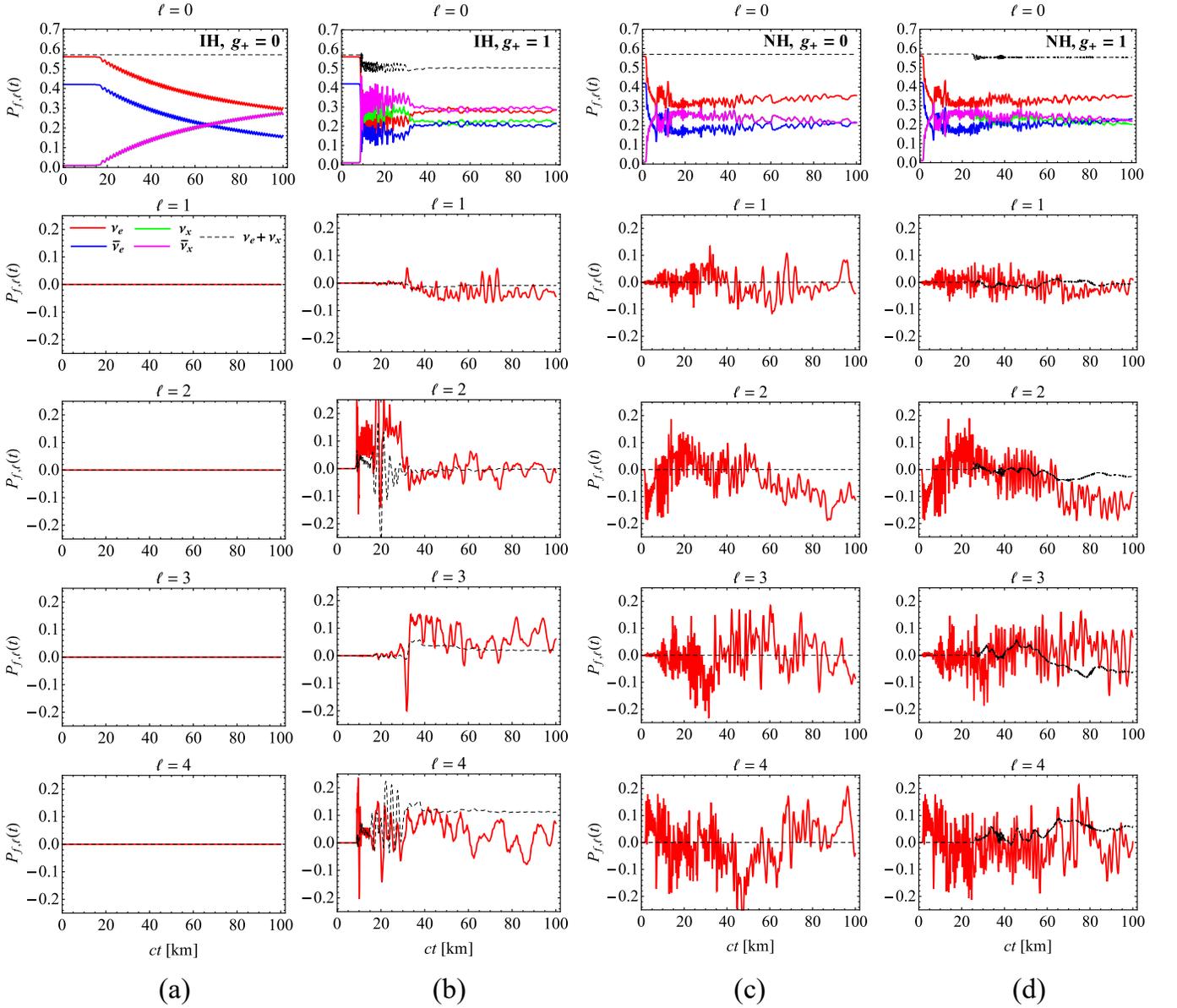}
        \end{center}
        \caption{Evolution of the Legendre harmonics $P_{f,l}(t)$ of the neutrino flavor probabilities for a decreasing neutrino density
        $n_\nu(t) = 3 \times 10^{29}~\text{cm}^{-3} / (1 + ct / (50~\text{km}))^2$ on top of a matter background
        with $n_{e} = 10^{29}~\text{cm}^{-3}$ and $n_n = 1.5 n_e$: (a) IH, $g_+ = 0$, (b) IH, $g_+ = 1$, (c) NH, $g_+ = 0$, (d) NH, $g_+ = 1$.
        The dashed line corresponds to the total $\nu_e + \nu_x$ probability.}
        \label{fig:MSW_effect}
    \end{figure}

    \subsection{Effect of background matter}
    \label{sec:Simulation_C}%

    Finally, we qualitatively analyze the impact of the MSW term on the development of instabilities, to complement the
    linear stability analysis (Fig.~\ref{fig:instabilityRates_mono_matter}). Namely, we use constant background electron
    and neutron densities $n_{e} = 10^{29}~\text{cm}^{-3}$, $n_n = 1.5 n_e$, and a time-dependent neutrino self-coupling
    $\mu(t) = G_{\text{F}} \sqrt2 n_\nu^{(0)} \times R_\nu^2 / (R_\nu + ct)^2$ with $n_\nu^{(0)} = 3 \times 10^{29}~\text{cm}^{-3}$.
    Such a setup mimics a decrease of the neutrino density with distance $r \equiv R_\nu + ct$ from the center of a supernova
    with a neutrino sphere radius $R_\nu = 50~\text{km}$ and virtually interpolates between the neutrino densities
    $3 \times 10^{29}~\text{cm}^{-3}$ at $ct = 0$ and $\sim 3 \times 10^{28}~\text{cm}^{-3}$ at $ct = 100~\text{km}$.
    Another advantage of this setup is that after the neutrino density has fallen down sufficiently, the oscillations
    get suppressed by the background matter effect, which prevents them from a chaotic late-time behavior observed within the constant-density setup.
    Indeed, Fig.~\ref{fig:MSW_effect}(b) reveals that in the case of inverted hierarchy, NSSIs drive the neutrino flavor
    probabilities to neat plateaus with a characteristic pattern $P_{e(x),0} \approx P_{\bar{x}(\bar{e}),0}$,
    and, as a result, a $\nu/\bar\nu$ ratio close to unity (an analogous behavior was observed earlier within the single-angle scheme \cite{Kharlanov2021}).
    Similarly to the constant-density case in Fig.~\ref{fig:instabilityDevelopment_harmonics}, early excitation of
    the even Legendre harmonics driven by the NSSIs
    further propagates to the odd ones, however, unlike that case, the harmonics of the total $\nu_e + \nu_x$ probability
    `freeze' roughly at the moment the matter potential starts to dominate the neutrino self-coupling. Of course, the
    `snapshots' of the $l \ne 0$ harmonics taken at this crossing point and propagated to the asymptotic values of $P_{e,l} + P_{x,l}$,
    are virtually random in different simulation runs, thus, the corresponding distortions of the asymptotic angular
    neutrino distribution can hardly be distinguished experimentally from a Poisson noise of neutrino events.
    However, the neutrino deficit relative to antineutrinos, emerging as a result
    of an SO(3) symmetry breaking, is observable if (pseudo)scalar neutrino interactions are present
    [see the $l = 0$ plot in Fig.~\ref{fig:MSW_effect}(b)]. In a purely V--A
    theory, matter effects totally hinder angular symmetry breaking and a different flavor pattern shows up
    in the isotropic harmonic [Fig.~\ref{fig:MSW_effect}(a)]. It is no surprise, in fact: when $g_+ = 0$,
    the density matrix is (almost) block-diagonal and its $l \ne 0$ harmonics are stable [cf.~Fig.~\ref{fig:instabilityRates_mono_matter}(e,~f)],
    thus, one can apply a transformation to a corotating frame in the flavor space to its isotropic harmonic \cite{Duan2006_CollOsc}.
    This transformation suppresses the effect of the matter potential, as long as it is much greater than the vacuum oscillation
    frequency, virtually, reducing the problem to that with a zero mixing angle. The latter, in turn, manifests itself in the suppressed oscillation
    amplitude of the $l = 0$ harmonic in Fig.~\ref{fig:MSW_effect}(a).

    For completeness, we also present the flavor evolution in the case of the normal mass hierarchy in
    Fig.~\ref{fig:MSW_effect}(c,~d). One observes that, even though the neutrino-antineutrino conversion takes place in the $g_+
    \ne 0$ case here, its effect on the zeroth harmonic is quite faint. The very onset of the neutrino-antineutrino instability
    in this case takes place considerably later, i.e., at lower densities, than in the inverted hierarchy [cf.
    Figs.~\ref{fig:MSW_effect}(d) and~\ref{fig:MSW_effect}(b)], which is in line with the absence of the high-density linear `instability pool'
    in the normal hierarchy [see Fig.~\ref{fig:instabilityRates_mono_matter}(c,~f)].

    \section{Discussion}
    \label{sec:Discussion}%

    In the previous sections we have discussed the unstable modes of a homogeneous isotropic Majorana neutrino gas with NSSI-enhanced
    interactions, limiting ourselves to the modes with translational and axial symmetry. Despite the simplicity of the setup,
    we have obtained a number of interesting results regarding the onset and further development of new instabilities
    that exist in the presence of (pseudo)scalar NSSIs. Let us briefly summarize the outcomes of our analysis, put
    them into context of the existing results, and discuss a possible outlook.

    First, it is worth reiterating that unlike the instabilities caused by the V--A interactions, the neutrino-antineutrino instability
    introduced by the $g_+$ (pseudo)scalar coupling does \emph{necessarily} break spherical symmetry: this follows immediately
    from the transformation law of the off-diagonal blocks of the neutrino density matrix, describing coherent mixing
    of the two helicities~\cite{Kharlanov2021}. Indeed, given a preferred direction along the $z$ axis, the characteristic shape of the
    neutrino-antineutrino instability is $\sin\vartheta$ [see Fig.~\ref{fig:instabilityDevelopment_theta}(b)],
    which is neither dipole ($\cos\vartheta$), nor isotropic (i.e., independent of $\vartheta$). Also, the NSSI and V--A instability
    branches affect different blocks of the density matrix, thus, they develop independently in the linear regime. The preferred $z$ direction
    is physically set by the magnetic field initiating the instability via an interaction with the transition magnetic moment
    of the neutrino. Interestingly, despite its anisotropic character, the neutrino-antineutrino mode has the same properties
    in the linear regime, as those identified in Ref.~\cite{Kharlanov2021} within a semi-qualitative analysis in the spirit
    of the single-angle scheme, up to a factor of $2/3$ (see Eq.~\eqref{secular_rho_diag_1}).

    Second, the properties of the neutrino-antineutrino instability strongly depend on the neutrino mass hierarchy: at least
    for homogeneous modes we have been studying here, the inverted hierarchy allows for a vast high-neutrino-density
    instability domain, a `pool' that extends to, in principle, arbitrarily small NSSI couplings $|g_+|$ (see Fig.~\ref{fig:instabilityRates_mono_largemu}).
    At the same time, the conventional dipole flavor instabilities are suppressed in this hierarchy [see
    Figs.~\ref{fig:instabilityRates_mono}(e),~\ref{fig:instabilityRates_mono_matter}(e) and Ref.~\cite{Duan2013_Angular}].
    As a result, the high-density regime of the inverted hierarchy allows for a new channel of SO(3) symmetry breaking via (pseudo)scalar
    neutrino self-interactions. This channel presents the only way to break the spherical symmetry in the high-density regime
    (retaining, however, the translational invariance), and could thus serve as a signature of BSM neutrino interactions.
    Indeed, going beyond the linearized equations of motion, one observes that the boundary of the `instability pool' also marks
    a boundary between the two phases of the fully-developed, `thermalized' chaotic state of the neutrino gas, which can be
    distinguished, e.g., by the average electron neutrino probabilities [see Fig.~\ref{fig:Pasymp}(a)]. In this figure, one can
    clearly identify the SO(3)-invariant phase for high neutrino densities and small $g_+$, the conventional, low-density V--A phase with
    a broken symmetry, and the NSSI-induced phase with a rotational symmetry broken for both the probabilities of the
    individual neutrino flavors and the total neutrino numbers of all flavors (i.e., $\nu_e + \nu_x$). In fact, the very presence of a
    steady state and the approach to it (see Fig.~\ref{fig:thermalization}) reveal the features that are quite similar
    to several other simulations of the nonlinear regime of collective oscillations
    \cite{Duan2019_NLM, Duan2020_NRM, Richers2021_ParticleInCell, Mirizzi2015_transInv, Bhattacharyya2020_LateTime,
    KharlanovGladchenko2021}.

    Third, though the `hidden-sector' $g_+$-induced instability needs a nonzero neutrino transition magnetic moment to be
    triggered, its suppressed values $\mu_{12} \sim 10^{-24}\mu_{\text{B}}$ allowed in the Standard Model
    \cite{Giunti2009_nuEMP, GIM} are quite sufficient to dramatically change the properties of the neutrino gas
    within several tens of kilometers. Much more important is the value of the $g_+$ coupling, but, as follows from Fig.~\ref{fig:Pasymp}(a,~c),
    sensitivities at the level of $|g_+| \sim 0.1$ can be achieved. Another remarkable property of the neutrino-antineutrino
    instabilities worth mentioning is that they can be enhanced by the background matter: for example, for the neutrino
    self-coupling $\mu = 50\omega$, the fastest instability rates are $\kappa_{\text{max}} \approx 1.4\omega$ in vacuum and
    $\kappa_{\text{max}} \approx 6.8\omega$ in matter with $G_{\text{F}}\sqrt2 n_e = 50 \omega$ and $n_n = 1.5 n_e$
    [cf. also Fig.~\ref{fig:instabilityRates_mono}(f) and Fig.~\ref{fig:instabilityRates_mono_matter}(f)]. Finally,
    in the presence of NSSIs, collective oscillations of neutrinos with a decreasing number density soon lead
    to a state with a sizeable relative deficit of neutrinos, because of their partial conversion to antineutrinos [Fig.~\ref{fig:MSW_effect}(b,~d)].
    Moreover, this effect can be observed in both mass hierarchies, probably in part because of the matter-induced instability enhancement.

    Regarding the potential sensitivity to the $g_+$ NSSI coupling, it is worth comparing our results with an analysis of
    scalar-mediated neutrino-neutrino interactions presented in Ref.~\cite{Shalgar2019_SecretInteractions}. Namely,
    in terms of a model with a Lagrangian $\Lagr = g (\nu_a^{\mathrm{T}} C \nu_a) \phi$ featuring a scalar $\phi$ with a mass $M$,
    our potential sensitivity $|g_+| \lesssim 0.1$ translates into a constraint $|g| \lesssim 10^{-6} M / \text{MeV}$, which
    could exclude a considerable part of the heavy-mediator area on the $(M, g)$ plane (see Ref.~\cite{Shalgar2019_SecretInteractions}
    for the corresponding diagram). Of course, such a constraint assumes a considerably heavy mediator,
    otherwise the very use of the effective four-fermion vertex~\eqref{L_nunu} is illegal. Thus, apart from this limitation, it
    appears that neutrino-antineutrino instabilities and conversions in supernova neutrinos could be a useful tool for probing
    (pseudo)scalar neutrino interactions, complementing other existing approaches.

    It is worth adding, as well, that the homogeneous instabilities discussed in the present study naturally require
    a further study within a broader context of instability \emph{waves}, i.e., perturbations proportional to~$e^{\ii(\Omega t - \bvec{Q}\cdot\bvec{x})}$.
    The resulting dispersion relations could change in part the conclusions drawn from the homogeneous ($\bvec{Q} = 0$) case,
    as is known within the V--A theory~\cite{Duan2015_NLM, Duan2019_DR}, but the status of the signature
    neutrino-to-antineutrino flavor transformations amplified by the NSSI-induced instabilities will probably stay the same.
    This, however, should constitute a subject of further studies.

    \section*{Acknowledgments}
    The author is grateful to Alexander Grigoriev for fruitful discussions. The problem discussed in the present paper
    is not to be confused with a similar but different setup constituting a Master's thesis of Chen Zekun under
    the supervision of the author.

    \appendix

    \section{Evolution equation in the Legendre basis}
    \label{app:LegendreBasis}
    We rewrite here the evolution equation~\eqref{rho_evolution} on the neutrino density matrix in the language of the
    Legendre decomposition~\eqref{rho_Legendre_def}. For the expansions which follow, let us quote the necessary properties
    of the Legendre orthogonal polynomials \cite{LegendrePolynomials}:
    \begin{gather}
        P_l(z) = \frac{1}{2^l l!}\frac{\diff^l}{\diff z^l} (z^2 - 1)^l, \quad
        \int_0^\pi P_l(\cos\vartheta) P_{l'}(\cos\vartheta) \cdot 2\pi\sin\vartheta\diff\vartheta = \frac{4\pi \delta_{l,l'}}{2 l + 1};
        \label{Pl_Rodrigues}
        \\
        P_0(\cos\vartheta) = 1, \quad P_1(\cos\vartheta) = \cos\vartheta, \quad P_l(\cos(\pi-\vartheta)) = (-1)^l P_l(\cos\vartheta);
        \label{P0_P1}
        \\
        \cos\vartheta P_l(\cos\vartheta) = \frac{l+1}{2l + 1} P_{l+1}(\cos\vartheta) + \frac{l}{2l + 1} P_{l-1}(\cos\vartheta).
        \label{Legendre_recurrence}
    \end{gather}
    Additionally, for a complex-valued function $f(\vartheta) = \frac{1}{4\pi} \sum_{l=0}^\infty f_l P_l(\cos\vartheta)$,
    the inverse transformation and the Parseval's theorem take the form:
    \begin{equation}
        f_l = (2l+1) \int_0^\pi f(\vartheta) P_l(\cos\vartheta) \cdot 2\pi\sin\vartheta\diff\vartheta,
        \quad
        \int_0^\pi |f(\vartheta)|^2 \cdot 2\pi\sin\vartheta\diff\vartheta = \frac{1}{4\pi} \sum_{l=0}^\infty \frac{|f_l|^2}{2l+1}.
    \end{equation}
    Finally, since our Hamiltonian~\eqref{rho_evolution} contains multiplication by $\sin\vartheta$, it is worth introducing
    its Legendre expansions:
    \begin{eqnarray}
        \sin\vartheta &=& \sum_{l=0}^\infty \varsigma_l P_l(\cos\vartheta),
        \quad
        \varsigma_l = \frac{2l+1}{4\pi} \int_0^\pi \sin\vartheta P_l(\cos\vartheta) \cdot 2\pi\sin\vartheta\diff\vartheta,
        \label{sine_Legendre_expansion}
        \\
        \sin\vartheta P_{l'}(\cos\vartheta) &=& \sum_{l=0}^\infty \varsigma_{ll'} P_l(\cos\vartheta), \quad
        \varsigma_{ll'} = \frac{2l+1}{4\pi} \int_0^\pi \sin\vartheta P_{l}(\cos\vartheta) P_{l'}(\cos\vartheta) \cdot 2\pi\sin\vartheta\diff\vartheta.
        \label{Legendre_sineMultiplication}
    \end{eqnarray}
    Because of the parity properties~\eqref{P0_P1}, $\varsigma_{2l+1} = 0$ and $\varsigma_{ll'} = 0$ unless $l$ and $l'$ have the same parity.
    The even coefficients $\varsigma_{2l}$ can be evaluated, e.g., by direct integration of Eq.~\eqref{sine_Legendre_expansion}
    using the explicit Rodrigues' representation~\eqref{Pl_Rodrigues} and an identity
    $\int_0^\pi \sin\vartheta \cdot \cos^{2k}\vartheta \sin\vartheta \diff\vartheta = \frac12 \sqrt\pi \Gamma(k+1/2) / \Gamma(k+2)$,
    which leads to
    \begin{equation}
        \varsigma_{2l} = \frac{(4l+1) \sqrt\pi}{4} \sum_{k=0}^{l} \frac{(-1)^{k + l} \Gamma(k + l + 1/2)}{k! (k+1)! (l-k)!}.
    \end{equation}
    The $\varsigma_{ll'}$ coefficients can be expressed in terms of $\varsigma_n$ by transforming a product of Legendre polynomials
    in the integrand of Eq.~\eqref{Legendre_sineMultiplication} into a sum of Legendre polynomials \cite{PP_identity}:
    \begin{gather}
        P_{l}(\cos\vartheta) P_{l'}(\cos\vartheta) = \sum_{k = 0}^{\min(l,l')}
            \frac{2(l+l'-2k) + 1}{2(l+l'-k) + 1} \frac{a_{k} a_{l-k} a_{l'-k}}{a_{l+l'-k}} P_{l+l'-2k}(\cos\vartheta),
        \quad a_k \equiv \frac{\Gamma(k+1/2)}{k! \sqrt\pi},
        \\
        \quad \varsigma_{ll'} = \sum_{k = 0}^{\min(l,l')} \frac{2l+1}{2(l+l'-k) + 1} \frac{a_{k} a_{l-k} a_{l'-k}}{a_{l+l'-k}}
        \varsigma_{l+l'-2k}.
    \end{gather}

    Now, on the right-hand side of the evolution equation~\eqref{rho_evolution}, commutation with a constant term
    $h_{\text{vac}} + h_{\text{mat}}$ in the Hamiltonian does not change the angular momentum $l$, while commutation with
    $h_{\text{AMM}} \propto \sin\vartheta$ leads to a contraction with the $\varsigma_{ll'}$ coefficient matrix.
    In the collective Hamiltonian~\eqref{h_self}, integration of
    $(1 - \cos\vartheta \cos\vartheta') \mathbb{K}(\varrho(\vartheta'))$ over $2\pi\sin\vartheta' \diff\vartheta'$ simply gives
    $\mathbb{K}(\varrho_0) - \cos\vartheta \mathbb{K}(\varrho_1) / 3$, while the NSSI term
    $\sin\vartheta \sin\vartheta' \mathbb{L}(\varrho(\vartheta'))$ yields
    $\sin\vartheta \sum_{n=0}^\infty \varsigma_{n} \mathbb{L}(\varrho_{n}) / (2n+1)$. Thus, the collective Hamiltonian
    is linear in the three functions $1, \cos\vartheta$, and $\sin\vartheta$; multiplication of these functions by the density
    matrix in $[h_{\text{self}}(\vartheta), \varrho(\vartheta)]$ can be readily done by virtue of
    Eqs.~\eqref{Legendre_recurrence}, \eqref{Legendre_sineMultiplication}. As a result, one arrives at the system~\eqref{rho_evolution_Legendre}
    governing the evolution of the density matrix harmonics $\varrho_l(t)$.

\end{document}